\DocumentMetadata{}

\documentclass[sigconf]{sty/acmart}
\clearpage{}\usepackage{colortbl}

\usepackage{breakurl}
\usepackage{fancyhdr}
\usepackage{epsfig}
\usepackage{balance}
\usepackage{color}
\usepackage{comment}
\usepackage{epic}
\usepackage{epsf}
\usepackage{listings}
\usepackage{makecell}
\usepackage{multirow}
\usepackage{textcomp}
\usepackage{xspace}
\usepackage{latexsym}
\usepackage[labelfont=bf,font=small,skip=5pt]{caption}
\usepackage{afterpage}
\usepackage{amstext}
\usepackage{amsmath}
\usepackage{mathrsfs}
\usepackage{graphicx}
\usepackage{tabularx}
\usepackage{tcolorbox}
\usepackage{threeparttable}
\usepackage{enumitem}

\usepackage{subfigure}

\usepackage{amsthm}

\usepackage{array}

\usepackage{longtable}
\usepackage{rotating}
\usepackage[ruled,linesnumbered]{algorithm2e}
\usepackage{fp}
\usepackage{siunitx}
\usepackage{xstring}

\usepackage{float}

\usepackage{ragged2e}

\usepackage{utfsym}
\usepackage{booktabs}

\usepackage{subcaption}
\usepackage{stfloats}

\usepackage{setspace}

\newtheorem{theorem}{Theorem}

\definecolor{bblue}{HTML}{4F81BD}
\definecolor{rred}{HTML}{C0504D}
\definecolor{ggreen}{HTML}{9BBB59}
\definecolor{ppurple}{HTML}{9F4C7C}
\definecolor{darkgray}{rgb}{0.66, 0.66, 0.66}
\definecolor{gray}{RGB}{136,136,136}
\definecolor{dkgreen}{rgb}{0,0.6,0}
\definecolor{gray}{rgb}{0.5,0.5,0.5}
\definecolor{mauve}{rgb}{0.58,0,0.82}
\definecolor{comment-red}{rgb}{0.8,0,0}

\newlength\tbspace
\newcolumntype{C}{c<{\hspace{\tbspace}}}

\SetCommentSty{mycommfont}

\makeatletter
\newcommand\footnoteref[1]{\protected@xdef\@thefnmark{\ref{#1}}\@footnotemark}
\makeatother
\newcommand{\DefMacro}[2]{\expandafter\newcommand\csname rmk-#1\endcsname{#2}}
\newcommand{\UseMacro}[1]{\csname rmk-#1\endcsname}

\usepackage{tikz}

\newcommand{\CodeIn}[1]{{\small \texttt{#1}}}
\newcommand{\Space}[1]{}

\makeatletter
\newcommand{\labitem}[2]{\def\@itemlabel{\textbf{#1}}
\item
\def\@currentlabel{#1}\label{#2}}

\newcommand{\RN}[1]{\textup{\uppercase\expandafter{\romannumeral#1}}}
\SetKwProg{Fn}{Function}{}{}

\newcommand{\PP}[1]{
\vspace{2px}
\noindent{\bf \IfEndWith{#1}{}{#1}{#1}}
}
\renewcommand{\paragraph}[1]{\smallskip\noindent\emph{#1}\quad}

\newcommand{\etal}{\textit{et al}.\xspace}

\newcommand{\Sys}{\mbox{\textsc{Nitro}}\xspace}
\newcommand{\SysAP}{\mbox{\textsc{Nitro's}}\xspace}
\newcommand{\SysR}{\mbox{\textsc{Nitro-R}}\xspace}

\newcommand{\ql}{QuickLog\xspace}
\newcommand{\qltwo}{QuickLog2\xspace}
\newcommand{\ed}{eAudit\xspace}
\newcommand{\edm}{eAudit-SEC\xspace}
\newcommand{\kl}{KennyLoggings\xspace}

\newcommand{\hardlog}{HardLog\xspace}
\newcommand{\nodrop}{NoDrop\xspace}
\newcommand{\omnilog}{OmniLog\xspace}
\newcommand{\omini}{OmniLog\xspace}

\newcommand{\tamperevident}{tamper-evident\xspace}
\newcommand{\loggingsystem}{logging system\xspace}

\newcommand{\ebpf}{eBPF\xspace}
\newcommand{\xlog}{XLog\xspace}

\newcommand{\sgx}{SGX-Log\xspace}
\newcommand{\custos}{Custos\xspace}
\newcommand{\stresstest}{stress-test\xspace}
\newcommand{\heading}[1]{\vspace{6pt}\noindent{\underline{\textsc{#1}}}}
\newcommand{\noskipheading}[1]{\noindent\textsc{\underline{#1}}}
\newcommand{\secref}[1]{Section~\ref{#1}}
\newcommand{\calK}{{\mathcal K}}
\newcommand{\Dom}{\mathrm{Dom}}
\newcommand{\bits}{\{0, 1\}}
\newcommand{\advA}{{\mathcal A}}

\newcommand{\muprf}{\mathrm{mu}\textrm{-}\mathrm{prf}}
\newcommand{\Adv}{\mathbf{Adv}}
\renewcommand{\Game}{\mathbf{G}}
\newcommand{\figref}[1]{Figure~\ref{#1}}

\DeclareMathAlphabet{\mathsfsl}{T1}{cmss}{m}{sl}
\DeclareMathAlphabet{\mathsc}{OT1}{cmr}{m}{sc}
\DeclareMathAlphabet{\mathsl}{OT1}{cmr}{m}{sl}

\newcommand{\gamesfontsize}{\small}

\newcommand{\twoColsNoDivide}[4]{
\begin{center}
        \framebox{
        \begin{tabular}{c@{\hspace*{.4em}}c@{\hspace*{.4em}}c}
        \begin{minipage}[t]{#1\textwidth}\setstretch{1.05}\gamesfontsize #3 \end{minipage}
        &
        \begin{minipage}[t]{#2\textwidth}\setstretch{1.05}\gamesfontsize #4 \end{minipage}
        \end{tabular}
        }
\end{center}
}

\newcommand{\getsr}{{\:{\leftarrow{\hspace*{-3pt}\raisebox{.75pt}{$\scriptscriptstyle\$$}}}\:}}
\newcommand{\Func}{\mathrm{Func}}
\newcommand{\FnO}{\textsc{Fn}}
\newcommand{\PROCEDURE}{\textbf{procedure} }

\newcommand{\ind}{\hspace*{10pt}}

\newcommand{\Range}{\mathrm{Rng}}
\newcommand{\setX}{{\mathcal X}}

\newcommand{\true}{\mathsl{true}}
\newcommand{\Update}{\mathrm{Update}}
\newcommand{\Sign}{\mathrm{Sign}}

\newcommand{\calS}{{\mathcal S}}
\newcommand{\faa}{\mathrm{fa}}
\newcommand{\FAtwo}{\mathrm{FA}}
\newcommand{\Agg}{\mathrm{Tag}}
\newcommand{\XLog}{\mathsf{XLog}}
\newcommand{\xor}{\oplus}
\newcommand{\advB}{{\mathcal B}}
\newcommand{\thref}[1]{Theorem~\ref{#1}}

\newcommand{\pad}{\mathrm{pad}}
\newcommand{\concat}{\|}

\newcommand{\Xs}{x_s}
\newcommand{\Xc}{x_b}
\newcommand{\Qs}{Q_s}
\newcommand{\Qc}{Q_b}
\newcommand{\Qi}{Q_i}
\newcommand{\Se}{S_e}
\newcommand{\Fs}{F_s}
\newcommand{\Fc}{F_b}
\newcommand{\Fi}{F_i}
\newcommand{\Fx}{F_k}
\newcommand{\Qx}{Q_k}

\newcolumntype{G}{>{\columncolor{gray!15}}c}
\newcolumntype{W}{>{\columncolor{white}}c}

\newcommand{\false}{\mathsl{false}}

\newcommand{\commentt}[1]{{\color{blue}{/\!/#1}}}

\newcommand{\forge}{\mathrm{forge}}
\newcommand{\AMAC}{\mathrm{MC}}
\newcommand{\combine}{\boxplus}
\newcommand{\combinv}{\boxminus}
\newcommand{\advD}{{\mathcal D}}
\newcommand{\propref}[1]{Proposition~\ref{#1}}
\newcommand{\Syss}{{\textsc{Nitro}}'s }
\newcommand{\ONew}{\mathsc{New}}
\clearpage{}

\begin{document}

\title{Rethinking Tamper-Evident Logging: A High-Performance, Co-Designed Auditing System}

\author{Rui Zhao}
\affiliation{\institution{University of Virginia}
  \city{Charlottesville}
  \country{USA}}
\email{dkw7xn@virginia.edu}

\author{Muhammad Shoaib}
\affiliation{\institution{University of Virginia}
  \city{Charlottesville}
  \country{USA}}
\email{ewe4gy@virginia.edu}

\author{Viet Tung Hoang}
\affiliation{\institution{Florida State University}
  \city{Tallahassee}
  \country{USA}}
\email{tvhoang@cs.fsu.edu}

\author{Wajih Ul Hassan}
\affiliation{\institution{University of Virginia}
  \city{Charlottesville}
  \country{USA}}
\email{hassan@virginia.edu} \begin{abstract}
    Existing tamper-evident logging systems suffer from  high overhead and severe data loss in high-load settings, 
		yet only provide coarse-grained tamper detection. 
		Moreover, installing such systems requires
		recompiling kernel code. To address these challenges, we present \Sys, a high-performance, tamper-evident  audit logging system
		that supports fine-grained detection of log tampering. 
		Even better, our system avoids kernel recompilation by using the eBPF technology.
    To formally justify the security of \Sys, we provide a new definitional framework for logging systems, 
		and give a practical cryptographic construction meeting this new goal. 
		Unlike prior work that focus  only on the cryptographic processing, we co-design the cryptographic part with the pre- and post-processing of the logs 
		to exploit all system-level optimizations. 
		Our evaluations demonstrate \Syss superior performance, achieving $10\times$-$25\times$ improvements in high-stress conditions and $2\times$-$10\times$ in real-world scenarios while maintaining near-zero data loss.
		We also provide an advanced variant, \SysR that introduces in-kernel log reduction techniques to reduce runtime overhead even further. 
\end{abstract}

 \maketitle

\section{Introduction}
\label{s:intro}

Audit logs are essential for tracking system activity and supporting security operations~\cite{custos,kbl+2017,king2003}. They capture detailed event sequences that help reconstruct attacks, identify stealthy behavior, and support root cause analysis, compliance auditing, and incident response~\cite{inam2022sok,priotracker2018,spade,king2003,lpm2015}. Their reliability makes them vital for both real-time detection and forensic investigation.

\smallskip
\noindent \textbf{Insecure Logging Systems}
\smallskip

\noindent
Current audit logging systems~\cite{camflow,lpm2015,linuxaudit,Sekar2023eAA,Tetragon,Sysdig,inam2022faust} predominantly trace and intercept system calls to capture log data. For example, Linux Auditd~\cite{linuxaudit}, the native Linux auditing tool, monitors system calls to record security-relevant events, but suffers from performance degradation under high workloads. To address these issues, Sekar \etal~\cite{Sekar2023eAA} introduced \ed, which uses eBPF technology~\cite{ebpf-bcc} within the Linux kernel to significantly improve audit logging performance. However, \ed\cite{Sekar2023eAA} and similar systems remain vulnerable to {\it race condition attacks} as demonstrated by Paccagnella \etal~\cite{kennylog}. These systems operate asynchronously, queuing system calls for later disk logging using a FIFO mechanism. This creates a tamper window, defined as the delay between when a system call occurs and when it is logged, which attackers can exploit to manipulate the log queue. Although \ed reduces this window, it does not {\it completely} eliminate it.

\smallskip
\PP{Tamper-Proof Logging Systems}
\smallskip

\noindent
To eliminate the tamper window, tamper-proof logging systems~\cite{nodrop,hardlog,omini} employ specialized hardware to make log modification infeasible. \nodrop~\cite{nodrop} uses Intel Memory Protection Keys (MPK)~\cite{Intel_MPK} to isolate memory regions containing logs, while \hardlog and \omini rely on external hardware devices and synchronous logging that pushes events to non-rewritable storage as they occur. These methods ensure strong integrity guarantees by preventing any post-generation log manipulation. However, this comes at a steep cost. Hardware-based isolation introduces substantial runtime overhead, especially under high throughput, as each log must be synchronously flushed before the system can proceed. Moreover, these systems require specialized hardware, making them impractical to deploy in general-purpose or cloud environments.

\smallskip
\PP{Tamper-Evident Logging Systems}
\smallskip

\noindent
As an alternative to costly hardware-based approaches, tamper-evident logging systems~\cite{kennylog,281386} offer software-based integrity guarantees and tackle race conditions attacks. These systems generate cryptographic tags for log entries, allowing auditors to detect tampering after the fact. For example, \kl~\cite{kennylog} generates an integrity tag per log using forward-secure MACs, while \qltwo~\cite{281386} improves verification efficiency by aggregating all per-log tags into a single value. 

\smallskip
\PP{Challenges in Tamper-Evident Logging Systems}
\smallskip

\noindent
Despite their advancements, tamper-evident logging systems face several challenges, which are described below.

\smallskip
\noindent {\it \textbf{C1:} Coarse-Grained Tamper Detection.}
Systems like \qltwo~\cite{281386} generate a single aggregate tag for an entire batch of logs, meaning that any tampering triggers a global verification failure without revealing which log was altered. While \kl~\cite{kennylog} improves detection granularity by tagging each log individually, it increases the storage overhead by 5\% and the running time by 40\%. 
Moreover, its security definition doesn't provide any theoretical guarantees for the extraction of unmodified logs.  

\smallskip
\noindent {\it \textbf{C2:} Insecure Tag Storage.}
Prior systems fail to ensure secure tag storage. In \kl~\cite{kennylog}, tags are stored in plaintext alongside logs, enabling truncation attacks where an adversary can delete recent logs and restore a previously stored tag to hide tampering. While \qltwo~\cite{281386} avoids storing tags altogether, this reduces auditability. 

\smallskip
\noindent {\it \textbf{C3:} Large Secret.} The cryptographic methods employed by tamper-evident systems~\cite{kennylog,281386} rely on secret keys and state information to ensure the integrity of the logs. However, keeping the size of those secrets small is crucial, as a larger secret state presents a greater risk of side-channel attacks. For instance, \kl~\cite{kennylog} precomputes about 3.2MB of secret keys, storing them in kernel space before system startup. 
While this approach accelerates the signing process, it is more vulnerable to side-channel attacks~\cite{281386}.

\smallskip
\noindent {\it \textbf{C4:} Difficult Maintenance.} Both \kl~\cite{kennylog} and \qltwo~\cite{281386} extend Auditd~\cite{linuxaudit} and require modifying and recompiling kernel modules to insert cryptographic hooks. These modifications introduce fragility into the kernel audit pipeline, risking crashes and making the systems error-prone to maintain across kernel updates.

\smallskip
\noindent {\it  \textbf{C5:} Excessive Overhead.} Existing loggers often introduce significant overhead, which can become intolerable under high system loads. In our experiments, we measured runtime overhead as the ratio of CPU time used by the logging systems to that of the benchmark itself.  \qltwo~\cite{281386} incurred a runtime overhead of 15.9\%--895.1\%. \qltwo outperforms \kl in their experiments, suggesting that \kl may impose even greater overhead.

\smallskip
\noindent {\it \textbf{C6:} Severe Data Loss.} Under high load, tamper-evident logging systems often drop large volumes of logs, compromising forensic analysis. This differs from race condition attacks, as it stems from overload rather than tampering. While acknowledged in non-secure loggers like \ed~\cite{Sekar2023eAA}, this issue is largely ignored in tamper-evident systems. Our experiments show that existing systems such as \kl~\cite{kennylog,281386} experience data loss rates of up to 98\%.

\smallskip
\noindent\textbf{Our Approach: \Sys}
\smallskip

\begin{figure}[!t]
	\centering
		\includegraphics[width=0.48\textwidth]{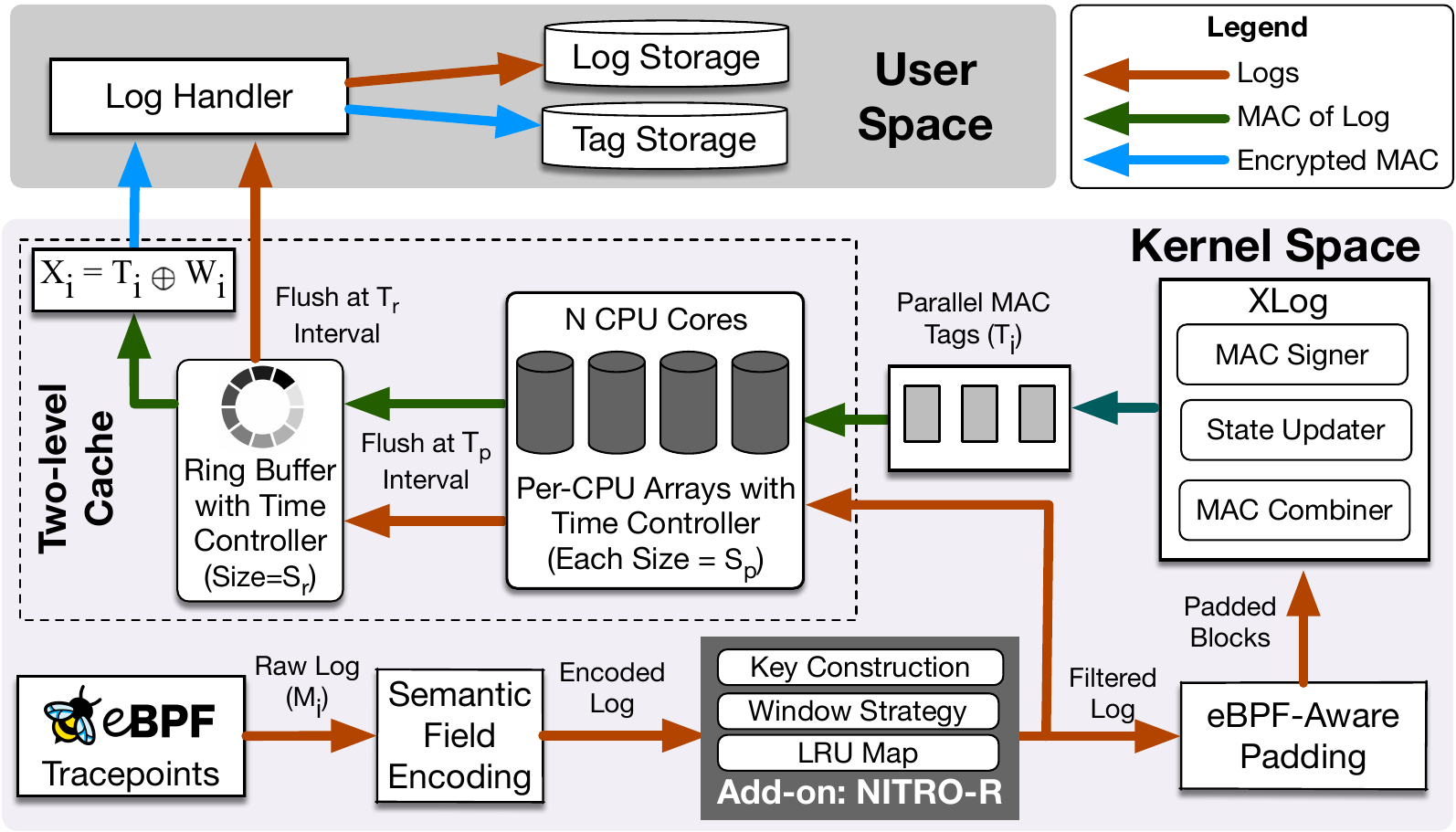}
	\caption{High-level workflow of \Sys.}
  \vspace{-3ex}
	\label{fig:arch:simplified}
\end{figure}

\noindent
To address the above-mentioned challenges, we present \Sys, a high-performance, tamper-evident audit logging system. To the best of our knowledge, \Sys is the first tamper-evident logging system to operate fully in eBPF. Figure~\ref{fig:arch:simplified} illustrates the overall architecture, and the main contributions are summarized below.

\heading{1. Provable security.}
To address C1 and C2, we extend the definitional frameworks in prior work~\cite{kennylog,281386}
to allow tags to be encrypted before being sent to storage. 
Our notion provides a unified syntax for \kl~\cite{kennylog} (where all tags are stored but not encrypted), 
\qltwo~\cite{281386} (where no tag is stored), and \Sys (where tags are \emph{occasionally} encrypted and stored). 
We formally show that our notion ensures that the extracted logs are unmodified.

Instead of building a logging system directly from a MAC like\qltwo~\cite{281386}, 
we define a new primitive that we call a \emph{MAC combiner}. 
This provides an abstraction on the tag aggregation process, and also a useful tool that 
may find applications beyond the context of logging systems. 
The idea of MAC combiner is inspired from the notion of aggregate MAC~\cite{XOR}, 
but the latter unfortunately does not work for the setting of logging systems.

\heading{2. eBPF-Compliant Implementation.}
To address C4, \Sys implements cryptographic logging entirely within the constraints of the eBPF runtime. These constraints include bounded stack sizes, static control flow, the absence of AES-NI~\cite{aesni} support, and the prohibition of dynamic loops. To navigate these limitations, we use Chaskey, an ISO-standard lightweight MAC, as our underlying MAC, selected for its compatibility with eBPF's verifier. Our contribution is not the use of Chaskey itself, but the careful
pre-processing of the logs before running Chaskey on them. The rationale for this design choice is further discussed in Section~\ref{s:design}.

 \heading{3. Parallel MAC Signing.}
A major cause of C5 is the bottleneck of centralized cryptographic state. 
To tackle this issue, \Sys assigns a separate signing context (tag, key, state) to each logical CPU core using eBPF’s Per-CPU Arrays. Each core signs its own logs independently, enabling decentralized MAC computation without synchronization overhead. While the parallel signing approach brings a big improvement on speed,
it worsens C3, because the state size is amplified by a factor of $N$, 
where $N$ is the number of logical cores. 
In our machine, $N = 36$, but we still manage to keep the state size at 1.2KB, which is much smaller than \kl~\cite{kennylog}. 

 \heading{4. Two-level Cache with Time-Aware Flow Control.} To overcome C6, \Sys introduces a novel caching strategy that reduces log loss during high load. It uses a two-level architecture with Per-CPU Arrays and ring buffer, managed by time-aware controllers that schedules log transfers based on system parameters such as core count, buffer saturation, and event rate. This adaptive mechanism minimizes I/O pressure and maintains reliable logging.\footnote{\ed~\cite{Sekar2023eAA} uses a two-level caching mechanism to reduce data loss, but its simple weight-based scheduling strategy leads to frequent data copying, resulting in higher I/O overhead as shown in our experiments.}

\heading{5. Co-Designed Logging and Cryptographic Stack.}  To resolve C5 and C6, \Sys avoids retrofitting cryptographic layers onto existing loggers like Auditd~\cite{linuxaudit} and instead builds a unified logging and MAC framework from scratch. This integration enables optimizations that reduce both computational and memory overhead. For example, \Sys introduces eBPF-aware padding that avoids runtime padding of all log messages by statically defining log structures. It also applies semantic field encoding, such as mapping syscall names to 32-bit integers, to shrink MAC input size. These design choices lower CPU cycles and simplify deployment by removing the need for kernel patches or hardware dependencies.

\smallskip
\noindent\textbf{Even Better: \SysR}
\smallskip

\noindent  Building on \Sys, \SysR is the first in-kernel, eBPF-based implementation of an audit log reduction technique inspired by Auditrim~\cite{auditrim}, significantly lowering I/O overhead by filtering redundant logs before they reach user space. Unlike prior user space approaches~\cite{auditrim,tang2018nodemerge,loggc,inam2022sok,hossain+depend}, \SysR operates entirely within the eBPF environment, avoiding added latency and complexity. Porting Auditrim to eBPF introduces key challenges: adapting time windows to system load, operating within eBPF’s restrictive execution model, and efficiently managing in-kernel data structures for log reduction. \SysR addresses these with an eBPF-based LRU hash map and verifier-safe logic for real-time, low-overhead log reduction. Applied before \SysAP \xlog, \SysR preserves log integrity while improving throughput and storage efficiency. Our goal is not to propose a new log reduction method but to demonstrate the feasibility of efficiently implementing existing techniques within the kernel using eBPF for improved performance.

\smallskip
\noindent\textbf{Evaluation Results}
\smallskip

\noindent
We evaluated \Sys against state-of-the-art logging systems using stress-test benchmarks (from \ed~\cite{Sekar2023eAA}) and real-world benchmarks (aligned with \hardlog~\cite{hardlog}). \Sys demonstrated a substantial performance advantage, with improvements of 10$\times$--25$\times$ on stress-test benchmarks and 2$\times$--10$\times$ on real-world benchmarks, while preserving tamper-evident properties. In terms of data loss, \Sys achieved near-zero loss, compared to other logging systems, which experienced 31\%--98\% data loss under identical conditions. Additionally, \SysR reduced runtime overhead by an average of 34\% beyond \Sys's performance. To understand how each optimization in \Sys contributes to its performance, we also conduct a factorial design analysis. To assess integration feasibility, we built \edm by embedding \xlog into \ed. \edm suffered from poor performance and significant data loss, confirming the need for architectural rethinking.

\PP{Availability} Our source code and benchmarks are available at \url{https://github.com/DART-Laboratory/Nitro}.

\section{Preliminaries}

\noskipheading{Notations.} 
By $\Func(\Dom, \Range)$ we denote the set of all functions $f: \Dom \to \Range$. 
We use~$\bot$  as a special symbol to denote rejection, and it is assumed to be outside~$\bits^*$. 
We write $X \concat Y$ to denote the concatenation of two bit strings~$X$ and~$Y$.

If $\setX$ is a finite set, we let $x \getsr \setX$ denote picking an element of~$\setX$ uniformly at random and assigning it to $x$. 
If~$A$ is an algorithm, we let $y\gets A(x_1,\ldots ; r)$ denote running~$A$ on inputs $x_1,\ldots$ and coins $r$, and assigning the output to~$y$. 
By $y\getsr A(x_1,\ldots)$ we denote picking~$r$ at random and letting $y\gets A(x_1,\ldots ; r)$. 
We write $\advA^f$ to indicate that adversary~$\advA$ has oracle access to a function $f$.

\heading{Games.} We use the game-playing framework of Bellare and Rogaway~\cite{EC:BelRog06}. 
(See \figref{fig:prf} for an example.) 
We write $\Game(\advA) \Rightarrow b$ to denote the event of running game $\Game$
with an adversary $\advA$ that results in $b$. 
We also write $\Game(\advA)$ to abbreviate $\Game(\advA) \Rightarrow \true$.

\heading{MAC.} A Message Authentication Code (MAC) scheme 
is a function $F: \calK \times \Dom \to \bits^\tau$. 
It takes as input a key $K \in \calK$ and a message $M \in \Dom$, 
and then deterministically produces a tag $T \gets F(K, M)$.

A standard security goal for a MAC scheme is  to be a pseudorandom function (PRF). 
Informally, this means that an efficient adversary can't tell the tags of its chosen messages from truly random strings. 
Traditionally, PRF security is defined for the single-user setting, 
but in our application, it is more convenient to use the multi-security PRF notion. 
Specifically, we define the advantage of an adversary $\advA$ 
in breaking the (multi-user) PRF security of $F$ as
\[ 
\Adv^{\muprf}_{F}(\advA) = 2 \cdot \Pr[\Game^{\muprf}_F(\advA)] - 1 \enspace, 
\] 
where game $\Game^{\muprf}_F(\advA)$ is defined in \figref{fig:prf}. 
If the adversary only attacks user $1$ then the game is in the single-user setting. 
It is well-known that one can generically reduce the multi-user PRF security to the single-user one
via a hybrid argument.

\begin{figure}[t]
\twoColsNoDivide{0.25}{0.15}
{
\underline{\textbf{Game} $\Game^{\muprf}_F(\advA)$} \\[2pt]
$b \getsr \bits$;~ $v \gets 0$ \\
$b' \getsr \advA^{\ONew, \FnO}$;~ Return $(b' = b)$ \\[4pt]
\underline{\PROCEDURE $\ONew()$} \\[2pt]
$v \gets v + 1$;~ $K_v \getsr \calK$ \\
$f_v \getsr \Func(\Dom, \bits^\tau)$ 
}
{
\underline{\PROCEDURE $\FnO(i, M)$} \\[2pt]
If $(b = 1)$ then \\
\ind Return $F(K_i, M)$ \\
Else return $f_i(M)$ 
}
\vspace{-1ex}
\caption{Game defining the (multi-user) PRF security of a function $F: \calK \times \Dom \to \bits^\tau$.}
\label{fig:prf}
\vspace{-2ex}
\end{figure} 
\section{Tamper-Evident Logging Systems}
\label{s:tma}

In this section, we provide a foundational treatment for tamper-evident logging systems. 
In particular, we give a more general syntax and  a stronger security definition (that are needed by our system) than prior work. 
Moreover, instead of building a logging system by aggregating tags from a MAC like~\cite{281386}, 
we abstract this via a new tool that we call \emph{MAC combiner}, making the construction conceptually simpler. 
Finally, we show how to build a practical logging system $\XLog$ meeting our notion.

\subsection{Defining Security For Logging Systems}

 \noskipheading{Threat model and assumptions.} Our threat model is similar to that of prior work~\cite{281386,kennylog}. 
In particular, we consider an adversary that mounts an attack to escalate privilege; 
the system calls of this attack are recorded in the logs. 
Once the adversary gains root access, it can modify the logs in both memory and storage to hide the traces of the attack.
Between the time that the logs of the attack are generated and the moment that the adversary has root privilege, 
there is a short window. This time gap is too short for the logs to be sent \emph{asynchronously} to the user space for cryptographic processing, 
as demonstrated empirically via the race attack in~\cite{kennylog}. Still, we assume that it is long enough for a quick, \emph{synchronous}
cryptographic processing. 

We assume that logs are periodically sent to a different, trusted machine for forensic analysis. 
For the auditor to detect the tampering of the logs, the logging system will include some short tags
as a proof of integrity. 
\kl~\cite{kennylog},  for example, uses a tag for every single log. 
\qltwo~\cite{281386} instead uses a single \emph{aggregate} tag for the entire collection of logs
to save storage cost and reduce running time. 
Instead of going to either extreme, our system \Sys still maintains an aggregate tag in the memory, 
but occasionally saves the current tag to the log storage. 
By doing so, we provide  finer-grained information about the tampering to the auditor than \qltwo, 
with only a small increase of storage cost and running time. 
Still, if we store the tags in the clear like \kl then this is vulnerable to a \emph{truncation attack}: 
an adversary can replace the current tag by an old tag from the storage and delete recent logs. 
As a result, the tags in the storage needs to be encrypted, 
and thus the syntax of our system deviates from prior work.

For the auditor to verify the integrity tags, we assume that initially the 
auditor and the logging system share a short secret state; the logging system will then update the state
immediately each time it signs a log message. 
We assume that if a variable is overwritten, one cannot recover the old value. 
We also assume that before gaining root access, the adversary can't retrieve information of secrets in kernel memory.

We say that the logs are \emph{unaltered}  if the attacker only modifies the logs created \emph{after} it gains root access, 
meaning the logs containing the information of the attack still remains intact. 
Our goal is to ensure that if logs are altered then the auditor should be able to detect it. 

\heading{Syntax.}
We now give a general syntax that allows a logging system to store encrypted tags. 
 A \emph{logging protocol} $\Pi$  consists of a pair
of deterministic algorithms $(\Update, \Sign)$ and is associated with a state space~$\calS$, key space $\calK$, 
and a tag length~$\tau$. 
Algorithm $\Update$ takes as input a state $S^*$ and a boolean $b$. 
If $b = \false$ then it outputs an updated state~$S$ and a key $K$. 
If $b = \true$, in addition to $(K, S)$, it outputs another string $X \in \bits^\tau$. 
We require that the output $(K, S)$ of $\Update$ is independent of the boolean~$b$. 
Algorithm $\Sign$ takes as input a key $K$, a message $M$, a prior tag $T \in \bits^\tau$, 
and produces an updated tag $T^* \in \bits^\tau$. 

\begin{itemize}[leftmargin=*]
\item Initially, 
the auditor samples a root state $S \getsr \calS$ and a key $K \getsr \calK$. 
He then stores $(K, S, T)$ in the server's memory, with $T \gets 0^\tau$.

\item Once the $i$-th log message~$M_i$ is available, 
the logging system first updates the tag $T \gets \Sign(K, M_i, T)$. 
If it doesn't want to store an encryption of the tag, 
it will update the key and state via $(K, S) \gets \Update(S, \false)$. 
Otherwise, it will run $(K, S, X) \gets \Update(S, \true)$ and 
encrypts the tag via $X \gets T \xor X$.
The choice of encrypting the tag or not does not affect the output $(K, S)$
and the subsequent tags.  
Note that the key and state are updated immediately after signing a new message. 
This ensures that in a race attack~\cite{kennylog}, 
by the time the adversary gains root access, the keys signing the logs of its attack are already overwritten with subsequent keys. 
\end{itemize}
The auditor is later given  messages $(M'_1, \ldots, M'_r)$ and an aggregate tag~$T'$. 
Using procedure $\Agg$ in \figref{fig:faa},  the auditor can derive the final tag~$T^*$
for $(M'_1, \ldots, M'_r)$.\footnote{The $\Agg$ procedure always calls  $\Sign(\cdot, \true)$, 
whereas the logging system may call  $\Sign(\cdot, b)$ for any $b \in \{\true, \false\}$.  
Still, since $\Sign$ and $\Update$ are deterministic, and since the key and state that $\Sign$ outputs are independent of $b$, 
if the logs are not modified then the output of $\Agg$ is consistent with what the logging system produces. 
} 
If the logs are altered then $T^* \ne T'$.

\smallskip 
The syntax above captures all existing logging systems. 
For example, in \kl~\cite{kennylog}, the string $X$ that $\Sign$ generates is simply~$0^\tau$, meaning that tags are stored in the clear instead of being encrypted. 
Moreover, \kl always calls $\Sign(\cdot, \true)$, meaning that all tags are stored. 
In \qltwo~\cite{281386}, one always calls $\Sign(\cdot, \false)$, meaning that there is no tag storage.

\heading{Defining security.}
For an adversary~$\advA$ attacking a logging protocol $\Pi$, 
we define its advantage in breaking the \emph{forward authenticity} ($\FAtwo$) of~$\Pi$ as
\[ 
\Adv^{\faa}_\Pi(\advA) = \Pr[\Game^{\faa}_\Pi(\advA)] \enspace, 
\] 
where game $\Game^{\faa}_\Pi(\advA)$ is defined in \figref{fig:faa}. 
Initially the adversary generates log messages $(M_1, \ldots, M_q)$ and an internal state~$\sigma$.\footnote{
In practice, those log messages are honestly generated by the logging system to record the syscalls. 
It can only help the adversary by letting it choose those log messages.}
The game then samples a state $S_0 \getsr \calS$ and a key $K_0 \getsr \calK$, 
and derive the tag of $(M_1, \ldots, M_i)$ for ever $i \leq q$, and encrypts it.  
The adversary is then given back its internal state $\sigma$, the final tag $T$, 
the encrypted tag $X_i$ of $(M_1, \ldots, M_{i})$,  
and the current key and state. 
Its job is to produce a forgery $(M'_1, \ldots, M'_r, T')$. 

Recall that the auditor will later derive the aggregate tag $ T^*$ 
of messages $(M'_1, \ldots, M'_r)$. 
The adversary wins the game if (1) $r < q$ or  $(M'_1, \ldots, M'_q) \ne (M_1, \ldots, M_q)$, meaning the logs are altered,  
and (2) $T' = T^*$, meaning that the auditor fails to detect that the logs are altered.

\begin{figure}[t!]
\twoColsNoDivide{0.43}{0.1}
{
\underline{\textbf{Game} $\Game^{\faa}_\Pi(\advA)$} \\[2pt]
$(M_1, \ldots, M_q, \sigma) \getsr \advA$;~ $S_0 \getsr \calS$;~ $K_0 \getsr \calK$\\
$(X_1, \ldots, X_{q}, T) \gets \Agg(M_1, \ldots, M_q)$ \\
$(M'_1, \ldots, M'_r, T') \getsr \advA(S, K, X_1, \ldots, X_{q}, T, \sigma)$ \\
\commentt{Attacker must alter messages to win} \\
If $(r \geq q) \wedge ((M_1, \ldots, M_q) =  (M'_1, \ldots, M'_q))$ then return $\false$ \\
$(X^*_1, \ldots, X^*_{r}, T^*) \gets \Agg(M'_1, \ldots, M'_r)$ \\
Return $(T^* = T')$  \\[2pt]
\commentt{Generate final aggregate tag and encrypt intermediate tags} \\
\underline{\PROCEDURE $\Agg(M_1, \ldots, M_v)$}\\[2pt]
$T_0 \gets 0^\tau$;~ $S \gets S_0$;~ $K \gets K_0$ \\
For $i \gets 1$ to $v$ do  \\
\ind  $T \gets \Sign(K, M_i, T)$ \\
\ind  $(K, S, X_i) \gets \Update(S, \true)$;~ $X_i \gets T \xor X_i$ \\
Return $(X_1, \ldots, X_{v}, T)$ 
}
{

}
\vspace{-1ex}
\caption{
Game defining the $\FAtwo$ security of a logging protocol~$\Pi$.  
}
\vspace{-2ex}
\label{fig:faa}
\end{figure}

In the definition above, the adversary is given all encrypted tags $(X_1, \ldots, X_{q})$. 
In our system, the adversary only has access to some of those, 
because we only encrypt and store tags occasionally. 
In \qltwo, the adversary is given no encrypted tag at all. 
Giving this additional information can only help the adversary.

\heading{Discussion.} It is instructive to see why \kl~\cite{kennylog} fails the definition above. 
Recall that \kl, all tags are stored in the clear, meaning that $X_i$ is the aggregate tag of $(M_1, \ldots, M_i)$. 
Given $(S, K, X_1, \ldots, X_{q}, T)$, the adversary can win the game by picking an arbitrary number $r < q$,
and then outputting $(M'_1, \ldots, M'_r) = (M_1, \ldots, M_r)$ and $T' = X_r$. 
In this case the messages are altered, 
and yet $T'$ will match the tag $T^*$ derived by running procedure $\Agg$ in \figref{fig:faa} on $(M'_1, \ldots, M'_r)$. 
As a result, the adversary wins with advantage $1$.

\heading{Extracting unmodified logs.} All prior work stops when the auditor realizes that the logs have been altered. 
But in practice, one wants to extract unmodified logs for forensic analysis. 
We now show how to do this. 

Suppose the auditor is given a list $L = \{(X'_j, j)\}$ of index $j$ and encrypted tag $X'_j$ for messages
$(M'_1, \ldots, M'_j)$. 
Recall that the auditor can compute a list of $(X^*_1, \ldots, X^*_r)$ of encrypted tags 
by running $\Agg$ on $(M'_1, \ldots, M'_r)$. 
Let $s$ be the largest index that $X'_s = X^*_s$; if there is no such index then we let $s = 0$. 
Suppose that the original logs are $(M_1, \ldots, M_q)$.
If the logs have been altered then we claim that the forward authenticity notion ensures that $M'_i = M_i$, for every $i \leq \min\{s, q\}$. 
To see why, fix an index $d$. 
\begin{itemize}[noitemsep, nosep, label=$\bullet$, leftmargin=*]
\item First consider the case $d \leq q$. 
Recall that the adversary $\advA$ is given the state $S_q$, the key $K_q$, the tag $T_q$, and the encrypted tags $(X_1, \ldots, X_q)$. 
If we instead give $(S_d, K_d, T_d, X_1, \ldots, X_d)$ to $\advA$, 
this can only help because $\advA$ can derive  $(K_q, S_q, T_q, \allowbreak X_{d + 1}, \ldots, X_q)$. 
Forward authenticity then ensures that the adversary cannot produce the correct aggregate tag $T'$ of another 
$(M'_1, \ldots, M'_d) \ne (M_1, \ldots, M_d)$. 
As a result, it also cannot produce the encrypted tag $X' = T' \xor (T_d \xor X_d)$ of $T'$. 
\item Suppose that $d > q$. 
Since the logs are altered, we must have $(M'_1, \ldots, M'_q) \ne (M_1, \ldots, M_q)$. 
Since the adversary knows $(S_q, K_q)$, it can compute the mask $W$ for the $d$-th tag. 
Forward authenticity ensures that given the aggregate tag $T$ of $(M_1, \ldots, M_q)$
and the list of encrypted tags $(X_1, \ldots, X_q)$, 
the adversary cannot produce $(M'_1, \ldots, M'_d)$ with a correct tag $T'$. 
As a result, it also cannot produce the encrypted tag $X' = T' \xor W$ of $T'$. 
\end{itemize}
Still, aiming for log extraction would amplify the advantage by a factor $q$.

\heading{Discussion.} Given that the adversary can modify all logs in the storage that haven't been sent to the auditor, one may question the significance of log extraction.
We argue that log extraction is meaningful if one couples
a tamper-evident logging system with some non-rewritable hardware like HardLog~\cite{hardlog}. 
That is, each log will be immediately signed but there will be some  bounded delay before it is written to the non-rewritable storage. 
In that case,  most logs created before the adversary gets root privilege would remain intact, and thus it is critical to extract them. 
\subsection{MAC Combiner}

In this section, we formalize a new primitive that we call \emph{MAC combiner}. 
It allows one to sign $q$ messages (each with a fresh key) and then combine the $q$ tags into a single, short tag. 
This tool gives an  abstraction of the XOR construction that \qltwo~\cite{281386} uses.

\heading{Definition.} An MAC combiner $\AMAC$ is a pair $(G, \combine)$, 
where $G: \calK \times \bits^* \to \bits^\tau$ is a MAC, and $\combine: \bits^\tau \times \bits^\tau \to \bits^\tau$
is an operator. 
To compute the aggregate tag $T_q$ of messages $(M_1, \ldots, M_q)$ for keys $(K_1, \ldots, K_q)$, 
one would compute $T_i \gets T_{i - 1} \combine G(K_i, M_i)$ for every $i \leq q$, where $T_0 = 0^\tau$.

We define the advantage of an adversary $\advA$ breaking the unforgeability security of $\AMAC$ as
\[ 
       \Adv^{\forge}_\AMAC(\advA) = \Pr[\Game^{\forge}_\AMAC(\advA)] \enspace, 
\] 
where game $\Game^{\forge}_\AMAC(\advA)$ is defined in \figref{fig:forge}. 
Here the (stateful) adversary first requests to sees the aggregate tag $T_q$ of messages $(M_1, \ldots, M_q)$. 
The parameter $q$ is called the \emph{message count} of~$\advA$. 
The adversary then produces forgery messages $(M'_1, \ldots, M'_r)$ and a forgery tag $T$, with $r \leq q$. 
It wins if the aggregate tag $T'_r$ of $(M'_1, \ldots, M'_r)$ under keys $(K_1, \ldots, K_r)$
is also $T$.

\begin{figure}[t]
\twoColsNoDivide{0.4}{0.1}
{
\underline{\textbf{Game} $\Game^{\forge}_{\AMAC}(\advA)$} \\[2pt]
$(M_1, \ldots, M_q, \sigma) \getsr \advA$;~ $T_0 \gets 0^\tau$ \\
For $i \gets 1$ to $q$ do  $K_i \getsr \calK$;~ $T_i \gets T_{i - 1} \combine G(K_i, M_i)$ \\
$(M'_1, \ldots, M'_r, T) \gets \advA(\sigma, T_q)$;~ $T'_0 \gets 0^\tau$ \\
For $i = 1$ to $r$ do $T'_i \gets T'_{i - 1} \combine G(K_i, M'_i)$ \\
Return $((M'_1, \ldots, M'_r) \ne (M_1, \ldots, M_q)) \wedge (T'_r = T)$ 
}
{

}
\vspace{-0.5ex}
\caption{Game defining unforgeability security of a MAC combiner $\AMAC = (G, \combine)$. }
\label{fig:forge}
\vspace{-2ex}
\end{figure}

\heading{Discussion.} Our notion is inspired by the definition of aggregate MAC of Katz and Lindell~\cite{XOR}
but the latter doesn't work in the setting of logging systems. 
Under the notion of aggregate MAC, the adversary $\advB$ is given \emph{individual} tags of some messages~$M_i$
and has to predict the aggregate tag of $(M'_1, \ldots, M'_r)$. 
In the forward security for logging systems, the attacker $\advA$ is instead given the aggregate tag of $(M_1, \ldots, M_q)$
and has to find the aggregate tag of $(M'_1, \ldots, M'_r)$. 
The reduction from forward security to aggregate MAC simply fails. 
Specifically, 
given $\advA$ that picks $r < q$ and $M'_i = M_i$ for every $i \leq r$, 
the only plausible way $\advB$ can provide $\advA$ with the aggregate tag of $(M_1, \ldots, M_q)$
is to obtain the individual tags $T_1, \ldots, T_q$ of all $M_1, \ldots, M_q$. 
But then what $\advB$ later obtains from $\advA$ is the aggregate tag of $(M_1, \ldots, M_r)$
that is useless because $\advB$ can generate this via $T_1 \boxplus \cdots \boxplus T_r$.

\heading{Security analysis.} 
We say that $\combine$ is \emph{commutative} if $X \combine Y = Y \combine X$. 
Moreover, $\combine$ is \emph{invertible} if there is an associated inverse operator $\combinv: \bits^\tau \times \bits^\tau \to \bits^\tau$
such that if $Z = X \combine Y$ then $Z \combinv Y = X$. 
Note that if $\combine$ is both commutative and invertible then for any fixed $X \in \bits^\tau$, if we pick $Y \getsr \bits^\tau$
then $X \combine Y$ is uniformly distributed over $\bits^\tau$.

The following result shows that if $G$ is a good PRF and $\combine$ is commutative and invertible then $\AMAC = (G, \combine)$ is a good MAC combiner. 
To have an intuition of why this is true, let's pretend that $G$ is a truly random function (instead of merely a PRF). 
Then given the aggregate tag $T$ of $(M_1, \ldots, M_q)$, the aggregate tag $T'$ of $(M'_1, \ldots, M'_r)$ is still uniformly distributed over $\bits^\tau$, 
thanks to the commutativity and invertibility of $\combine$. 
Hence the chance that the adversary can guess $T'$ is about $2^{-\tau}$. 
The proof is in Appendix~\ref{A:agg_security_proof}.

\begin{proposition}
Let $\AMAC = (G, \combine)$ be a MAC combiner of tag length~$\tau$, 
where $\combine$ is commutative and invertible. 
Then for any adversary $\advA$ whose message count is at most $q$, 
we can construct an adversary $\advB$  of about the same running time such that
\[ 
\Adv^{\forge}_\AMAC(\advA) \leq \Adv^{\muprf}_G(\advB) +  \frac{1}{2^\tau}\enspace. 
\] 
Adversary $\advB$ makes at most $2q$ queries, with two queries per user.  
\label{P:agg}
\end{proposition}

\heading{Instantiation.} A simple choice for $\combine$ is the xor operator, which is used in the aggregate MAC construction of Katz and Lindell~\cite{XOR}. 
This is the choice in the \qltwo logging system~\cite{281386}, and also our system. 
But this is not the only possible choice. One can use, for example, modular addition, 
but this is less efficient.

\subsection{The XLog Construction}

\heading{The XLog construction.} 
We now show how to build a logging system $\XLog[\AMAC, F]$ from a MAC combiner $\AMAC = (G, \combine)$ 
where the key length and tag length are both $n$ bits, and a PRF $F: \bits^n \times \{0, 1, 2\} \to \bits^n$. 
In practice $n$ should be 128. 
The specification of $\XLog$ is given in \figref{fig:xlog}.

\heading{Security analysis.}  
The following result shows that if $\combine$ is invertible then $\XLog$ provides forward authenticity; 
the proof is in Appendix~\ref{A:XLog_security_proof}. 

\begin{figure}[t]
\twoColsNoDivide{0.4}{0.05}
{
\underline{\PROCEDURE $\XLog[\AMAC, F].\Update(S, b)$} \\[2pt]
$S' \gets F_S(0)$;~ $K' \gets F_S(1)$; $X \gets F_S(2)$ \\
If $b$ then return $(K', S')$ else return $(K', S', X)$ \\[4pt]
\underline{\PROCEDURE $\XLog[\AMAC, F].\Sign(K, M, T)$} \\[2pt]
$T' \gets T \combine G_K(M)$;~ Return $T'$
}
{
}
\caption{The $\XLog[\AMAC, F]$ logging system, with $\AMAC = (G, \combine)$.}
\label{fig:xlog}
\vspace{-2ex}
\end{figure}

\begin{theorem}
Let $F: \bits^n \times \bits^* \to \bits^n$ be a PRF, and let $\AMAC = (G, \combine)$ be a MAC combiner
of $n$-bit key and $n$-bit tag where $\combine$ is invertible. 
For an adversary~$\advA$ making $q$ signing messages, we can construct adversaries $\advB$ and $\advD$ of about the same running time such~that
\[ 
\Adv^{\faa}_{\XLog[\AMAC, F]}(\advA) \leq \Adv^{\muprf}_F(\advB) + \Adv^{\forge}_{\AMAC}(\advD) \enspace. 
\] 
Adversary $\advB$ makes $3q$ oracle queries, with exactly 3 queries per user. 
The message count of $\advD$ is also $q$.   
\label{T:XLog_security}
\end{theorem}

\noskipheading{Proof ideas.} Let $r$ be the number of forgery messages that $\advA$ outputs, and $q$ be the number of signing messages. 
To reduce the forward authenticity of $\XLog$ to the unforgeability of $\AMAC$, we need to model the keys as independent, uniformly random strings. 
Unfortunately, if $r > q$ then one can't treat the keys $K_{q + 1}, K_{q + 2}, \ldots$ as independent, 
because they can be derived from the state $S_q$, and this state is given to the adversary. 
Our proof gets around this issue by exploiting the fact that $\combine$ is invertible, 
which allows us to reduce the case $r > q$ to the case $r = q$. 

Assume that $r \leq q$. Since we derive the keys, states, and task masks via a PRF, 
one can view those as independent, uniformly random strings. 
As a result, the key, state, and encrypted tags that the adversary $\advA$ receives are independent of the prior keys, states, and tags, 
and thus can be ignored. 
In this simplified view, the forward authenticity game becomes the unforgeability game, 
meaning that the security of $\AMAC$ implies the security of $\XLog$.  \section{\Sys}
\label{s:design}
Building on the theoretical foundation of \xlog and the MAC combiner, we now present \Sys, a practical logging system designed for real-world deployment without kernel modification. In this section, we discuss how the cryptographic guarantees of \xlog are integrated into the constrained execution environment of eBPF, addressing various performance bottlenecks through strategic architectural optimizations.

\subsection{Overview of eBPF \& Its Limitations}

The Extended Berkeley Packet Filter (eBPF)~\cite{ebpf_overview, ebpf-faster2, ebpf-faster3} is a revolutionary technology that enables user-defined programs to be safely executed within the operating system kernel. Developed from the Berkeley Packet Filter (BPF)~\cite{linux_bpf}, eBPF significantly broadens the scope of BPF beyond networking. It now supports general-purpose system tracing, security monitoring, and performance analysis by allowing programs to attach to various kernel hooks such as tracepoints, kprobes, and network events without requiring kernel modifications or additional modules~\cite{rice2023learning}.

\begin{figure}[!t]
	\centering
		\includegraphics[width=0.45\textwidth]{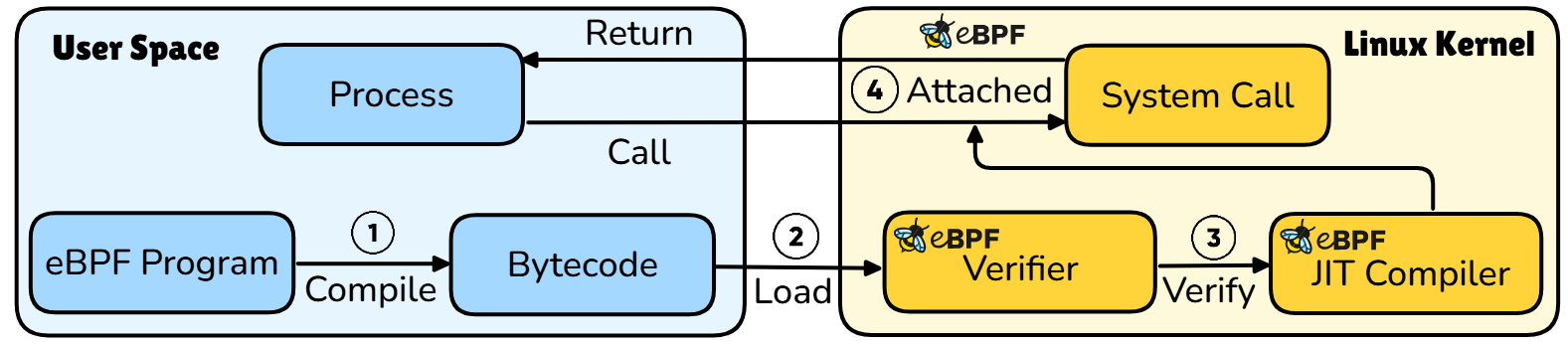}
	\caption{Workflow of \ebpf program compilation, verification, and attachment to system call hooks in the Linux kernel.}
  \vspace{-3ex}
	\label{fig:arch:ebpf}
\end{figure}

To ensure safety and stability, eBPF introduces a static verification engine that analyzes each program before loading, disallowing unsafe operations such as unbounded loops, invalid memory access, or kernel crashes. Once verified, eBPF programs are compiled using a Just-In-Time (JIT) compiler to run with near-native efficiency, while maintaining strict isolation from core kernel components.
Figure~\ref{fig:arch:ebpf} shows how an eBPF program is prepared and executed. It is first compiled in user space into bytecode, then passed to the kernel where it goes through a verifier for safety checks. If accepted, the bytecode is JIT-compiled into native machine code and attached to a system call. The hook is triggered when the corresponding system call is invoked. It can inspect both the call and its return value, and even block the system call if necessary.
For efficient communication between kernel and user space, eBPF provides high-performance data transfer mechanisms including perf buffers and ring buffer. Ring buffer is now widely recommended due to its lower latency and better multi-core scalability~\cite{ring}. Consequently, \Sys adopts ring buffer as the foundation of its logging pipeline.

Integrating existing logging systems, such as \qltwo~\cite{281386}, into eBPF initially appears straightforward, but practical obstacles render this approach infeasible. eBPF imposes several stringent limitations: it prohibits dynamic loops (L1), severely restricts memory usage (L2), rejects programs with high computational complexity (L3), and lacks support for hardware acceleration such as AES-NI~\cite{aesni} (L4). These constraints render standard cryptographic constructions impractical. For example, implementing AES without AES-NI leads to severe performance penalties, and memory limits prevent the inclusion of essential components like permutation tables. Exceeding these storage constraints results in memory allocation failures, making direct integration impractical.

\subsection{Selecting a Suitable PRF for eBPF}
Given eBPF's strict limitations, selecting an appropriate pseudo-random function (PRF) was essential. We required a PRF lightweight enough to fit eBPF's constraints, avoiding AES-based constructions and adhering to verifier limitations. 
The chosen PRF should have 128-bit output because $\XLog$ needs to use it  to derive 128-bit keys. 
These constraints ruled out candidates like SipHash~\cite{SipHash} due to their 64-bit output limitation. After evaluating alternatives, we selected Chaskey~\cite{Chaskey}, an ISO-standard PRF that provides provable security, significant performance advantages (approximately 8 faster than AES without AES-NI), and a compact, modifiable codebase suitable for broad industrial and governmental adoption. While Chaskey operates sequentially, which might suggest potential advantages in parallel designs, our compact encoding drastically reduces log message sizes (e.g., down to 24 bytes on average), diminishing any parallelization benefits in this context and emphasizing simplicity and efficiency.

\subsection{Integrating \xlog with eBPF: Challenges \& Optimizations}
\label{d:naive}
We start with a naive implementation of the blueprint above,
by adding $\XLog$ on top of eAudit~\cite{Sekar2023eAA}, with Chaskey as the underlying PRF.
We keep the architecture of eAudit, but remove its weighting mechanism.
This mechanism aims to classify logs based on their priority,
so that eAudit can send critical logs to the storage (that is \emph{assumed} to be non-rewritable in~\cite{Sekar2023eAA}) as quickly as possible.
In our case, the security model is different; the adversary may be able to write to the log storage after it gets root privilege. Hence
the weighting mechanism is useless in our threat model and can be dropped. We refer to this extended version of \ed as \edm.

However, evaluation (detailed further in \secref{s:eval}) revealed significant performance drawbacks with this naive approach. Specifically, integrating \xlog increased runtime overhead by approximately 5.8$\times$ and data loss by about 35$\times$. These outcomes underscored that \ed was fundamentally unsuitable for pairing with tamper-evident logging systems.

To address these challenges, we systematically optimized the architecture by exploiting \xlog’s design properties and adapting key aspects of the logging pipeline. We identified two core inefficiencies in \edm and redesigned the system accordingly:

\begin{itemize}[leftmargin=*]
  \item \textit{Streamlined MAC Computation with Parallel Signing:}
  In the naive approach, logs are retrieved entirely from memory before being processed by the MAC, introducing unnecessary latency. To mitigate this, we directly integrate eBPF calls into Chaskey's processing pipeline, building log message blocks incrementally and immediately performing MAC operations on each block. Additionally, the original approach utilizes a centralized signing key updated sequentially after each signing operation, severely limiting parallelism during system peaks. To alleviate this bottleneck, we implement a distributed signing mechanism by maintaining individual signing keys per CPU logical core using Per-CPU Array structures, dramatically reducing contention and increasing throughput.

  \item \textit{Fewer I/O requests:} In \ed~\cite{Sekar2023eAA}, the storage is assumed to be non-rewritable,
  so one needs to push log messages to the user space frequently to minimize the tampering window.
  In contrast, in our setting, the adversary can write to the storage after it gets root privilege,
  and thus having the logs in the storage or in memory make no difference in terms of security.
  Hence in our case, once a log is signed and the aggregate tag is updated,
  the log can continue to reside in the memory as long as space permits.
  We therefore can push the logs to the user space with much less frequency.

\end{itemize}

These optimizations must work within the limitations imposed by eBPF. Specifically, we adopt a set of verifier-compatible implementation techniques to handle its limitations. For L1, we combine a fixed-loop structure with a padding strategy (see Section~\ref{s:sign}) to safely process variable-length inputs. For L2, we minimize memory usage through compact log structures and avoid large buffers or dynamic allocation. For L3, we restructure the MAC computation using pointer-based staged processing to stay within the verifier’s instruction limits. Lastly, for L4, our lightweight PRF Chaskey well-suited for software-only execution. These design and implementation choices together enable efficient and secure integration of \xlog into the constrained eBPF environment.

\subsection{Parallel MAC Signing} \label{s:sign}
\noskipheading{How to implement signing.} 
Under Chaskey,  a message needs to be parsed into a sequence $M_1 \cdots M_m$, where each $|M_i| = 128$. 
(If the  message is fragmentary, padding is needed for the last block.)  
Initially, the tag~$X$ is initialized to $0^{128}$. 
One then iteratively goes over each message block~$M_i$ and updates $X$. 
Instead of running Chaskey on top of eAudit, 
we directly make eBPF calls to derive a message
and run Chaskey processing on each 128-bit block of the message on the fly.
Compared to the naive approach in \secref{d:naive},  
our approach hides the memory cost of retrieving the log message from memory, 
and also saves some cost due to type casting. 
Still, implementing this in the eBPF framework is tricky since only static loops are allowed (L1).

To get around the limitation above, we view a message as a pair $(P_1, P_2)$, 
where the first part $P_1$ consists of fixed-length arguments, such as the syscall name or PID, 
and the second part~$P_2$ consists of dynamic-length arguments. 
Let $\pad(P_1)$ be the string obtained by appending $0$'s to $P_1$
until the next multiple of 128 bits. 
We run a static loop to build  the message blocks of $\pad(P_1)$
and implement Chaskey's processing to derive an intermediate tag $X$. 
We then run another static loop; the number of iteration is a 
constant big enough to exceed the number of blocks in~$P_2$. 
In this loop, we resume Chaskey's processing with the tag~$X$ above. This novel strategy, which we refer to as \emph{eBPF-aware padding}, enables
MAC signing in eBPF by carefully structuring the message layout to comply with verifier constraints
while minimizing runtime costs. Conceptually, the process above is to run Chaskey on the string
$\pad(P_1) \concat P_2$. Since the length of~$P_1$ is uniquely determined by the syscall name,
the string $\pad(P_1) \concat P_2$ is a unique encoding of $(P_1, P_2)$. 

 At the first glance, one may wonder why we sign a longer message $\pad(P_1) \concat P_2$
instead of $P_1 \concat P_2$. 
While the latter option may save some cryptographic cost, it also adds 
overheads because  (1) $P_1$ and~$P_2$ are built from different data types, and some type casting is needed, 
and (2) this may cause memory misalignment
if the length of~$P_1$ is not a multiple of 4 bytes.

\heading{Semantic Field Encoding.} We designed a compact encoding of the logs that significantly reduces the log size similar to ~\cite{Sekar2023eAA,hossain2017sleuth}.
For example, in the Postmark benchmark, the average size of logs generated by the Linux daemon auditd
is 850B, whereas ours is only 24B.
The benefit of this log compression is even more important to our setting than that of eAudit~\cite{Sekar2023eAA},
because smaller logs means less cryptographic~cost.

\heading{Multiple signing lines.} If we have a single aggregate tag for all the logs, 
this creates a bottleneck during peak time because we can only use a single CPU core
to sign the logs one by one. 
A natural solution is to maintain multiple tags to enable parallel processing. 
This approach however has never been considered in the past, 
because it is unclear how to use multiple cores to sign logs concurrently without resorting to expensive multi-threading mechanisms. 
To deal with this issue, we leverage the Per-CPU Array data structure to have one tuple of (tag, key, state) per CPU logical core. 
Each CPU core only signs logs generated by that core and updates its own tag. 

Our approach parallelizes the cryptographic cost for both the server and the auditor. 
This performance gain however comes with a cost for security. 
First, the security definition in \figref{fig:faa} only considers a single user, 
but here we have~$N$ users, where~$N$ is the number of logical CPU cores. 
Single-user security still implies multi-user security via a standard hybrid argument, 
but concretely, the advantage is amplified by a factor of $N$. 
Moreover, instead of keeping just 32-byte secret in the memory (16B for the key and 16B for the state), 
we now have to keep $32N$ bytes of secret, which increases the risk of a side-channel attack. 
Still, in practice~$N$ is small. 
For example, in our machine, $N = 36$, 
and thus we keep about 1.2KB secret in memory. 
In contrast, KennyLoggings~\cite{kennylog} has 3.2MB secret. 

\subsection{Two-level Cache Design} \label{s:cache}

In eAudit~\cite{Sekar2023eAA}, the Per-CPU Array pushes data to the ring buffer at a maximum interval of $2^{24}$ ns,
and the ring buffer moves data to the user space after every $8$ operations.
That is, there are at least 8 transfers from the ring buffer to the user space per second.
This frequency is unnecessary in our setting, because once a log is signed and the key is updated,
the log can safely reside in the memory as long as space permits.
We instead use a two-level time controller to reduce the I/O cost.

\begin{itemize}[leftmargin=*,topsep=0pt]
\item \textit{Per-CPU Array $\rightarrow$ ring buffer:} The Per-CPU Array would push data to the ring buffer periodically,
for every $T_p$ seconds.
\item \textit{Ring buffer $\rightarrow$ user space:} The ring buffer moves data to the user space
after every $T_r$ seconds.
\end{itemize}

\heading{How to Choose Optimal Parameters}
\label{s:model}
We provide a flexible parameter setting mechanism in \Sys. Here, we discuss how to choose the optimal parameters as a reference.
The runtime performance and reliability of our system are mainly controlled by four key parameters: the Per-CPU Array size \(S_p\), the ring buffer size \(S_r\), the first-level flush interval \(T_p\), and the second-level flush interval \(T_r\). To understand the trade-off between runtime overhead and data loss, we simulate all parameter combinations across a large design space. The evaluation is conducted on all seven stress-test benchmarks, each running with 16 threads, and we report the average results. Specifically, we sweep \(S_p\) from 0.5KB to 32KB, \(S_r\) from 1MB to 64MB in exponential steps, \(T_p\) from 5ms to 300ms in linear steps of 50ms, and \(T_r\) from 500ms to 3000ms in steps of 500ms. This results in a total of 1,764 configurations, each representing a candidate deployment strategy.

\begin{figure}[!t]
	\centering
		\includegraphics[width=0.34\textwidth]{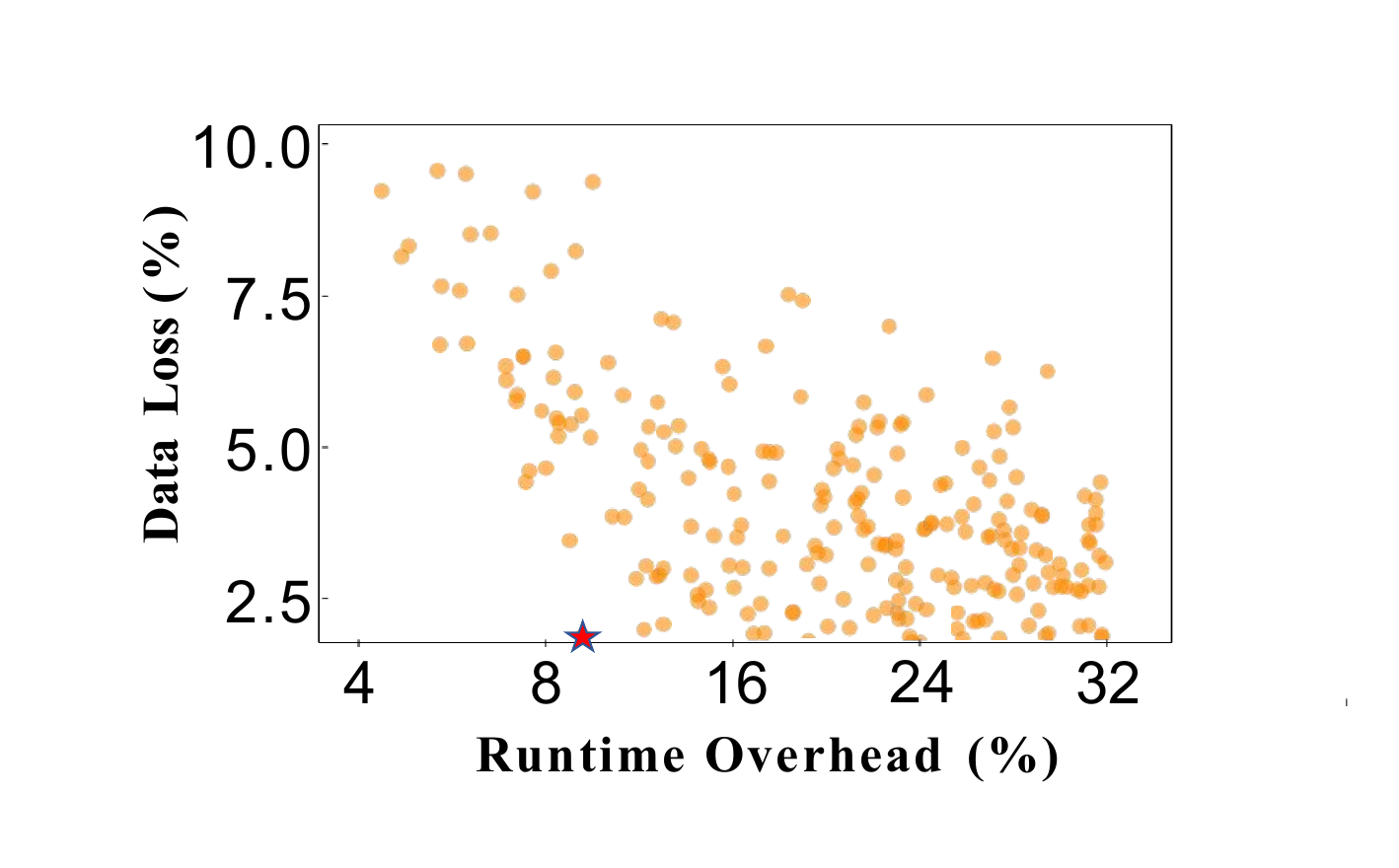}
	\caption{Trade-off between data loss and overhead under different parameter settings on seven stress-test benchmarks. The red pentagram marks the selected optimal configuration.}
  \vspace{-3ex}
	\label{fig:overhead_vs_data_loss}
\end{figure}

Our results, presented in Figure~\ref{fig:overhead_vs_data_loss}, include only the configurations that keep data loss below 10\% and overhead below 32\% to avoid visual clutter and highlight practical deployment choices. We observe that overly aggressive flushing (i.e., small \(T_p\) and \(T_r\)) can significantly reduce data loss but at the cost of high I/O overhead. Conversely, extremely lazy flushing schedules (e.g., large \(T_r\) or small \(S_r\)) may keep overhead low but risk frequent buffer overflows. In particular, we identify that settings with \(S_p \geq 8\text{KB}\), \(S_r \geq 16\text{MB}\), \(T_p \in [55\text{ms}, 255\text{ms}]\), and \(T_r \in [1000\text{ms}, 2000\text{ms}]\) offer a favorable balance between runtime overhead and data loss.

We pick the Per-CPU Array size \(S_p\) to be 32KB, which is empirically enough to avoid queue overflow in the Per-CPU Array. To avoid queue overflow in the ring buffer, if we have \(N\) CPU logical cores and the ring buffer size is \(S_r\), we need \(S_r \geq T_r \cdot (N \cdot S_p) / T_p\). In our system, \(N = 36\), and we pick \(T_p = 200\,\text{ms}\), \(T_r = 1\,\text{s}\), and \(S_r = 64\,\text{MB}\), which also fall within the optimal range identified in our experiments. Compared to eAudit-SEC, our system reduces the number of transfers from the ring buffer to user space by at least \(8\times\), and reduces the frequency of transfers from the Per-CPU Array to the ring buffer by approximately \(12\times\).

\subsection{Log Integrity Checking}

Once the log entries $(M'_1, \ldots, M'_r)$ and tags are transferred to user space, they are stored in their respective storages, as illustrated in Figure~\ref{fig:arch:simplified}. Later, during audit, the auditor can recompute a final tag~$T^*$ over $(M'_1, \ldots, M'_r)$ using the procedure $\Agg$ shown in \figref{fig:faa}. If the logs have been tampered with, the recomputed tag will not match the original. Unlike prior work~\cite{kennylog}, we consider a stronger adversary who may arbitrarily modify the logs stored on a compromised machine. This setting reflects more realistic threat scenarios but also introduces new challenges for verifying log integrity. In particular, the attacker may attempt to truncate logs or insert forged entries without detection.
To address this, \xlog provides formal guarantees that such tampering will be detected during audit. Moreover, the auditor can still identify all unaltered logs before the point of modification. These properties are formally proven in Section~\ref{s:tma} under the security model of \(\mathsf{XLog}\).

\subsection{Putting It All Together}

Creating and managing logs consists of three steps: (i) pre-processing (e.g., monitoring events to create logs and encoding them), (ii) cryptographic processing (e.g., signing logs with a MAC), and (iii) post-processing (e.g., caching the logs). Prior works either study step (ii) and use Auditd for the other steps or focus on steps (i) and (iii) without incorporating (ii) into their design. This results in a disconnection: one cannot combine existing state-of-the-art solutions such as \qltwo and \ed. Our work adopts a holistic yet modular approach, addressing all three components. We first establish the theory for step (ii) in \secref{s:tma}, independent of steps (i) and (iii). Then, in this section (\secref{s:design}), we co-design steps (i) and (iii) and select an instantiation for step (ii).

\subsection{Security Analysis}
Section~\ref{s:tma} provides the formal proof. Here, we follow the analysis style of prior work~\cite{kennylog}.
We examine key attack scenarios and explain how \Sys offers full or partial protection.

\begin{itemize}[leftmargin=*]
\item \textbf{A1: Log Truncation Attack.} An adversary attempts to remove or truncate part of the system logs to hide malicious activities or erase forensic evidence.
\item \textbf{A2: Side-Channel Attack.} An adversary performs side-channel attacks targeting kernel memory to extract the tag generation keys, which could enable the forgery of logs and subsequent tags.
\item \textbf{A3: Replay Attack.} An adversary reuses previously captured valid log messages to deceive the system or forge historical events, potentially leading to incorrect analysis or audit results.
\end{itemize}

{\bf A1} poses a threat to all \tamperevident systems that do not encrypt tags, in cases where the system outputs tags together with plaintext logs. In such systems, the adversary can truncate the log by removing the later part that contains malicious activities. The remaining earlier logs and their corresponding tags remain consistent, allowing the adversary to bypass detection.
\Sys addresses this threat by introducing tag encryption. In \Sys, the output tag is encrypted. Without the correct round key and state, the adversary cannot recover the original tag value from the encrypted tag. Additionally, \Sys's key update and tag aggregation mechanisms ensure each tag reflects the latest log content, preventing mismatches. As defined in Section~\ref{s:tma}, \Sys fully prevents log truncation attacks.

{\bf A2} is partially mitigated in \Sys through key rotation and per-core logging, which significantly reduce side-channel attack surfaces. Unlike prior work~\cite{kennylog} that stores megabytes of precomputed secrets in memory, \Sys requires only 48 bytes of secret material per core, including the key, state, and tag. This design supports parallel signing and removes synchronization overhead. The use of per-core registers in modern CPUs and frequent access to secrets during MAC computation lower the risk of cache-based leakage. Per-core isolation also ensures that compromising one core does not affect the security of others.

{\bf A3} is fully prevented in \Sys through its forward-secure design. Even if an adversary gains access to the log storage device along with plaintext logs and encrypted tags, tampering with existing logs or forging new entries is infeasible. This is because each log and its tag are derived from prior logs, per-round secret keys, and an evolving state, rather than generated independently. Without access to the initial state and keys, an adversary cannot perform a successful replay attack, even with full knowledge of the cryptographic algorithms.
 \section{\SysR}
\label{s:reduction}

Prior work~\cite{inam2022sok,priotracker2018,hossain+depend,auditrim,swift2020,loggc,elise} shows that system logs often contain redundant entries that inflate storage and processing overhead without improving detection or forensic value. Auditrim~\cite{auditrim} demonstrated that suppressing such duplicates is both feasible and effective. Inspired by these findings, we present \SysR, an in-kernel mechanism that eliminates duplicate log entries in real time \textit{before} any cryptographic processing.

\SysR is the first log reduction system implemented entirely within the kernel using eBPF. Unlike previous systems~\cite{kcal,tang2018nodemerge,loggc,inam2022sok,hossain+depend} that require kernel modifications or rely on post-processing in user space, \SysR performs safe, real-time log reduction directly inside the kernel within the strict constraints of the eBPF verifier. This design required addressing several non-trivial technical challenges:

\begin{itemize}[leftmargin=*]

\item First, eBPF imposes strict restrictions such as prohibiting unbounded loops, dynamic memory allocation, and complex control flow. To comply with these limitations, we designed a verifier-safe log reduction logic using only static memory access patterns, bounded loops, and minimal branching, ensuring real-time performance without verifier rejections.

\item Duplicate detection requires efficient identification of repeated log entries based on their semantic content. \SysR addresses this by constructing a compact, structured key composed of the syscall name, process ID, and arguments, and mapping it to timestamps using an in-kernel LRU hash table. This enables precise and efficient duplicate detection under tight memory and instruction constraints.

\item The effectiveness of log reduction depends on an appropriate time window for determining whether two events are duplicates. \SysR supports dynamic adjustment of this time window, computed as $\min(1, 2 \times (t_1 - t_0) + 0.001)$. Here, $t_0$ and $t_1$ represent the previous and current timestamps of the same event. While our evaluation uses a fixed value of $T_W = 1$ second as suggested by Auditrim~\cite{auditrim}, the system is capable of adapting to changing system conditions.

\item Continuous log generation poses risks of unbounded memory growth. \SysR mitigates this by leveraging the built-in eviction capabilities of eBPF’s LRU hash map, ensuring bounded memory usage and maintaining long-term scalability without requiring kernel modification.

\end{itemize}

Crucially, \SysR performs log reduction before any cryptographic signing by \xlog, ensuring that only unique log entries are authenticated. This ordering guarantees the integrity and authenticity of final logs while significantly reducing both MAC computation and I/O overhead. Moreover, \SysR is designed to be modular. Although we implement a time-window-based log reduction inspired by Auditrim, the system can easily be extended to support alternative log reduction techniques such as semantic filtering, priority-aware sampling, or event coalescing~\cite{tang2018nodemerge,loggc,inam2022sok}, as long as they can be implemented under eBPF's programming model. The detailed algorithm is provided in Algorithm~\ref{alg:SysR} (Appendix), which processes each log entry in real time, identifies near-duplicates using structured keys and timestamps, and emits only unique entries to the reduced log buffer. \section{Evaluation}
\label{s:eval}

In this section, we evaluate the performance and reliability of \Sys. In particular, we investigate the following research questions (RQs):

\begin{itemize}

    \item[\textit{RQ1}] What is the runtime overhead of \Sys compared to state-of-the-art logging systems?
		(Sections~\ref{sub:micro_eval})
    
    \item[\textit{RQ2}] How much does \Sys reduce data loss during high system loads compared to state-of-the-art systems? (Section~\ref{sub:data_loss})
    
    \item[\textit{RQ3}] How much parallel signing and two-level cache improve \Sys’s performance and reduce data loss? (Section~\ref{sub:opt_perf})

    \item[\textit{RQ4}] How does \SysR perform in terms of runtime overhead and data loss? (Section~\ref{sub:cmp_sysr})
    
    \item[\textit{RQ5}] What is the runtime memory usage of \Sys? (Appendix~\ref{sec:memory_usage})

    \item[\textit{RQ6}] How do \Sys and \nodrop compare in terms of runtime overhead?\footnote{Since \nodrop~\cite{nodrop} requires Intel-specific hardware, we conducted its experiments separately on a machine equipped with the necessary Intel chips. To avoid skewing the comparisons in RQ1 and RQ2, we isolate this evaluation in RQ6, where we directly compare \Sys and \nodrop on various benchmarks.} (Appendix~\ref{sec:nodrop-cmp})

    \item[\textit{RQ7}] How well does \Sys detect log tampering and how quickly does it respond under real-world attacks? (Appendix~\ref{app:real-world-attacks})
    
    \item[\textit{RQ8}] Do the logs generated by \Sys and \SysR maintain correctness under causality analysis principles? (Appendix~\ref{app:log-correctness})

\end{itemize}

\heading{Implementation} We implement \Sys via the BPF Compiler Collection (BCC) version 0.29.1~\cite{ebpf-bcc}. We select 68 critical syscalls as our monitoring targets; see Appendix~\ref{app:syscalls} for the list of those. To monitor system calls, we use APIs from BCC like  \texttt{TRACEPOINT\_PROBE} to listen for syscall invocations, and use \texttt{args} to get parameters of each syscall~\cite{ebpf-bcc-doc}.  We implement the $\XLog$ construction in C.
The signing of a message involves two static for-loops;
they are manually unrolled for performance improvement.
Informed by \thref{T:XLog_security}, we pick the tag length to be 64 bits, and represent each tag as an unsigned 64-bit integer.
Each log would contain an additional 32-bit identifier for the CPU core. This identifier is just a bookkeeping and is \emph{not} processed by the MAC. The log handler is written in Python 3. It receives data from kernel mode through \texttt{bpf.ring\_buffer\_poll}~\cite{ebpf-bcc-doc},
then parses and stores them on disk.

\heading{Evaluation Setup} Our experiments used VMs on an AMD EPYC 9654 server. Each VM was configured with 300 GB RAM, 200 GB SSD, and 36 logical CPU cores, running Ubuntu 22.04 (kernel 6.5.0). Only essential software was installed to minimize interference, and each experiment was repeated 30 times.
In all experiments, the relative standard deviation ranged from 3.91\% to 14.05\%, ensuring result reliability.

\heading{Benchmarks Selection Criteria.} To comprehensively evaluate \Sys’s runtime overhead, we used a diverse set of benchmarks to compare it against state-of-the-art logging systems~\cite{281386,Sekar2023eAA,nodrop}. For high-load conditions, we adopted all seven benchmarks from \ed~\cite{Sekar2023eAA}, referred to as \textit{Stress-test Benchmarks}, including Postmark (mail server simulation) and Linux kernel compilation to represent resource-intensive workloads. To evaluate performance in practical settings, we incorporated benchmarks from \hardlog~\cite{hardlog}, labeled as \textit{Real-world Benchmarks}, which include Firefox (web browsing) tested with Speedometer~\cite{firebench}, OpenSSL for cryptographic operations~\cite{openssl}, and lmbench~\cite{lmbench}. Further details on both sets of benchmarks are provided in Tables~\ref{tab:micro} and~\ref{tab:real} in Appendix~\ref{app:benchmarks}. Although \qltwo~\cite{281386} shares similar evaluation goals with \ed~\cite{Sekar2023eAA}, we excluded its benchmarks for two reasons: (1) its syscall-level tests (e.g., open, write) generate significantly lower workloads than our stress-test benchmarks, and (2) its application-level benchmarks overlap with those already included in our real-world evaluation. Consistent with prior work in secure audit logging~\cite{hardlog,kennylog,281386,omini}, we did not include DARPA and ATLAS datasets~\cite{darpaoptc,darpatc,atlas}, as they are not designed to assess the performance of audit logging systems.

\heading{Baseline Selection.} 
We selected state-of-the-art systems for the evaluation of \Sys. \qltwo, the best-performing tamper-evident system, was chosen over \ql and \kl due to its superior performance. For tamper-proof systems, we selected \nodrop~\cite{nodrop}, which offers strong performance and accessible hardware requirements, excluding \omnilog~\cite{omini} and \hardlog~\cite{hardlog} due to their stricter hardware demands. Prior evaluations confirm that \nodrop outperforms both alternatives~\cite{nodrop}, making it an optimal baseline. For non-secure systems, we included \ed, which provides the best balance of runtime overhead and data loss, excluding other non-secure tools~\cite{Tetragon, Sysdig, tracee2, linuxaudit, pasquier2015:camflow-tcc} as \ed already outperforms them~\cite{Sekar2023eAA}. To assess the feasibility of integrating cryptographic protections into \ed, we implemented \edm, which incorporates \Sys’s cryptographic method (\xlog) within \ed’s architecture.\footnote{We did not attempt to combine \ed with \qltwo, as \qltwo depends on AES-NI hardware acceleration, which is incompatible with eBPF due to its restrictive verifier and lack of hardware instruction support.} We also excluded existing log reduction systems~\cite{inam2022sok,Xu:2016,elise,tang2018nodemerge,hossain+depend,ding2023case}, as our goal is not to propose a new log reduction scheme but to demonstrate that existing techniques can be adapted for in-kernel processing via eBPF.

\heading{Evaluation Metrics.} We define runtime overhead \( O_{\text{runtime}} = (T_{\text{total}} - T_{\text{benchmark}}) / T_{\text{benchmark}} \), where \( T_{\text{benchmark}} \) is the CPU time used by the benchmark without logging and \( T_{\text{total}} \) is the time with logging enabled. This metric follows the methodology in \ed and quantifies the additional overhead introduced by the logging system. We define data loss as \( P_{\text{loss}} = D_{\text{discarded}} / D_{\text{total}} \), where \( D_{\text{discarded}} \) is the amount of log data dropped during operation and \( D_{\text{total}} \) is the total generated log data. This measures the system's ability to capture logs reliably.

\begin{figure*}[t!]
  \centering
  \includegraphics[width=0.92\textwidth]{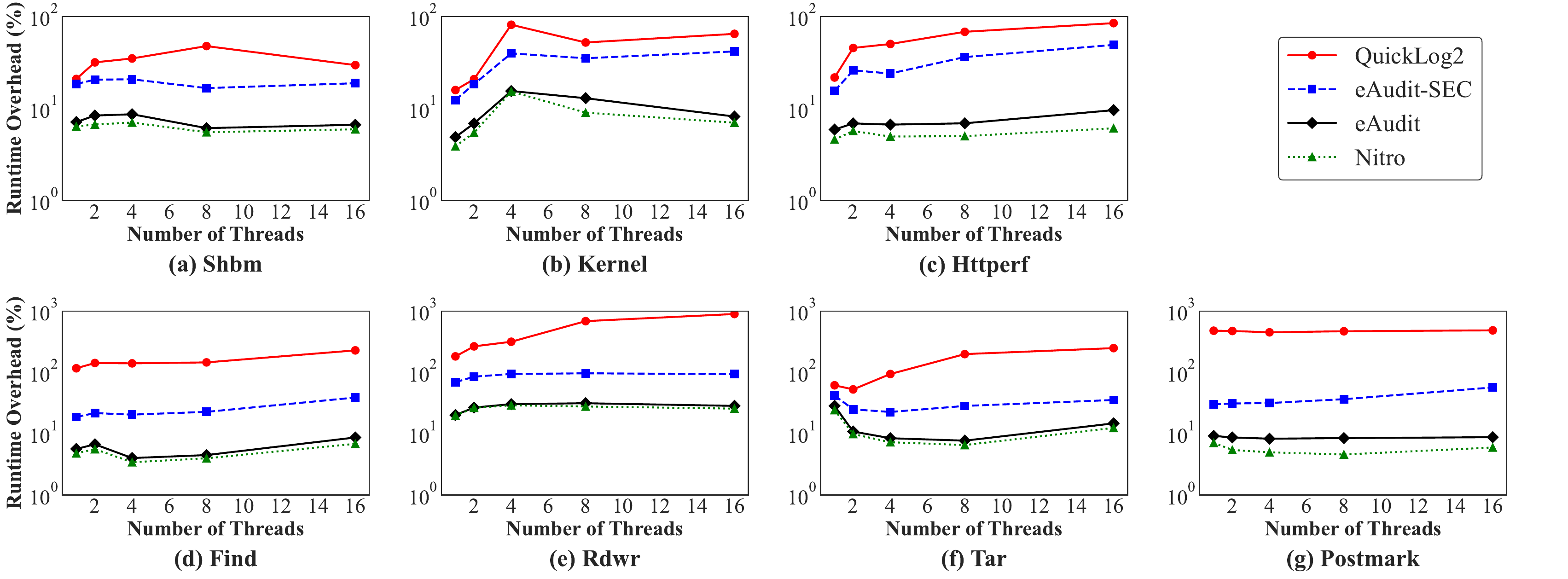}
  \caption{Runtime overhead comparison under the stress-test benchmark.}
  \label{fig:micro_mark}
  \vspace{-1ex}
\end{figure*}

\begin{figure}[t!]
  \centering
  \includegraphics[width=0.46\textwidth]{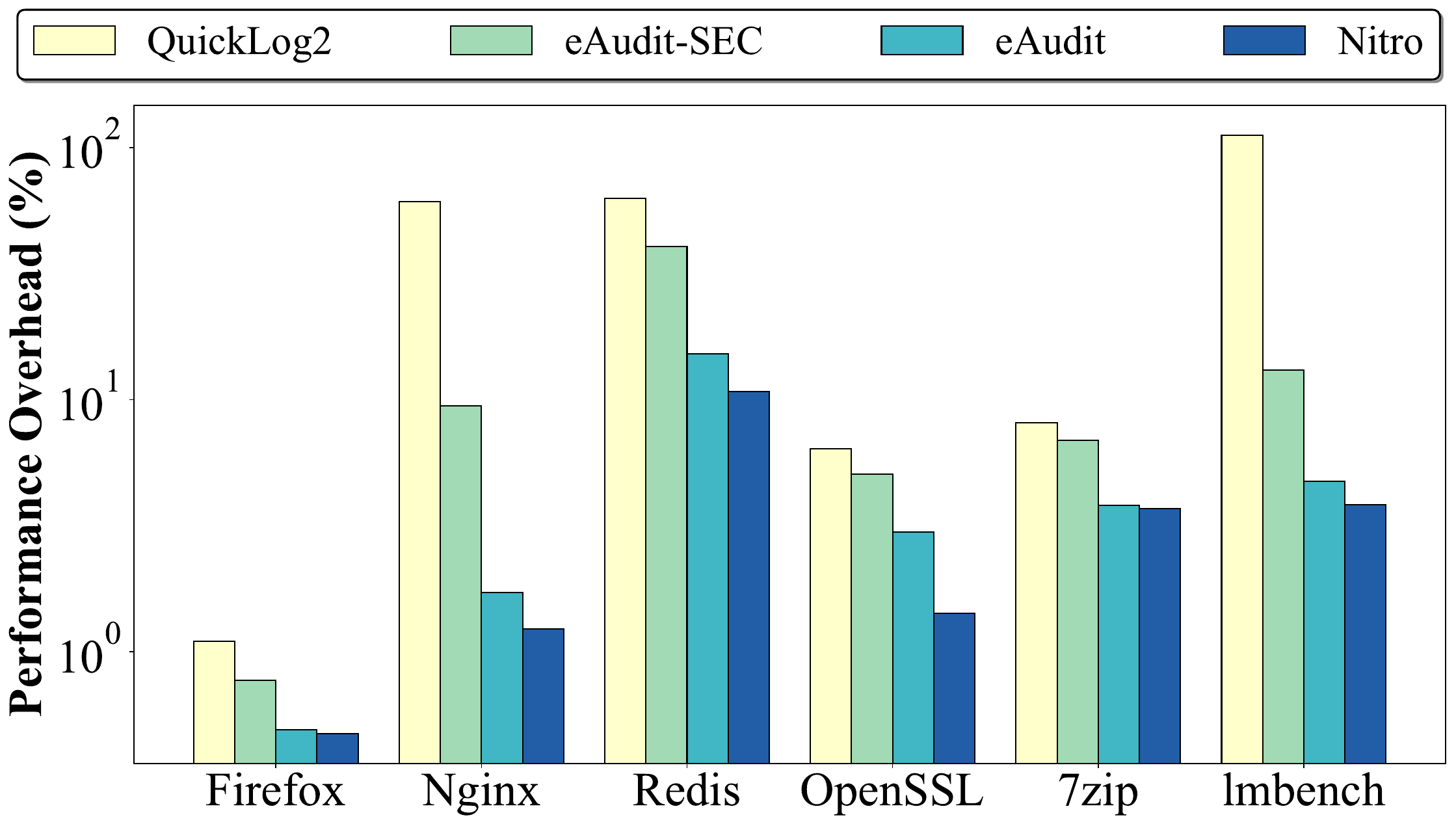}
  \caption{Runtime overheads for the real-world benchmark.}
  \label{fig:lmbench}
  \vspace{-3ex}
\end{figure}

\subsection{RQ1: Runtime Overhead}
\label{sub:micro_eval}

\heading{Stress-test Benchmarks.}
Figure~\ref{fig:micro_mark} shows that \Sys consistently outperforms \qltwo across all seven stress-test benchmarks. During repeated high-load tests, particularly with PostMark~\cite{postmark}, systems running \qltwo frequently crashed. This instability stems from modifications to the core logging system (Linux Auditd~\cite{linuxaudit}), which drops logs under load, leading to conflicts with the MAC logic. In contrast, \Sys avoids these issues entirely and offers superior stability. For I/O-intensive tasks such as \CodeIn{find} and \CodeIn{rdwr}, \Sys avoids overhead spikes and outperforms \qltwo by up to 30$\times$ while maintaining consistent performance. \Sys also achieves runtime overhead comparable to or lower than \ed, owing to its optimized \xlog and reduced kernel I/O. This confirms that strong security and low overhead can be achieved together. Results further show that simply adding cryptographic protections to \ed (\edm) is insufficient. We applied MACs to \ed using Chaskey~\cite{Chaskey}, as in \Sys. Although \edm benefits from eBPF efficiency and occasionally outperforms \qltwo, it fails to achieve both speed and security.

\heading{Real-world Benchmarks.}
\label{real_mark}
Figure~\ref{fig:lmbench} shows that \Sys consistently achieves lower runtime overhead than all other systems across a diverse set of real-world benchmarks, demonstrating its superior runtime efficiency. This result highlights \Sys’s ability to sustain high performance under practical conditions where system demands vary. While performance differences are generally less pronounced in lower-load scenarios, \Sys still maintains an average performance advantage exceeding 4$\times$ over \qltwo, reflecting the benefits of its architecture and MAC optimizations. These findings confirm \SysAP ability to deliver minimal overhead under a wide range of workloads and establish it as a high-performance, secure logging solution that addresses key limitations of legacy \loggingsystem-based architectures.

These results underscore a central insight: retrofitting cryptographic protections onto legacy loggers like Auditd, as in \qltwo, leads to instability and inefficiency due to design mismatches. In contrast, \Sys co-designs logging and cryptographic components from the ground up. Compact encoding, eBPF-aware padding, verifier-safe key management, and per-CPU MACs ensure compatibility with eBPF while reducing log size, CPU usage, and I/O pressure. These choices enable \Sys to achieve fast logging.

\subsection{RQ2: Data Loss}
\label{sub:data_loss}

In each experiment, data loss is calculated as the ratio of the average number of lost logs (across $n = 30$ independent runs) to the average total number of logs.
The ratio of two normal variables is not normally distributed, and thus one cannot use the standard way to compute confidence intervals. 
Instead, we employ Fieller's method~\cite{Fielle} to compute confidence intervals for the ratio of two normal variables. 

In particular, consider two normal variables $X$ and $Y$ in $n$ observations with sample mean $\mu_X$ and $\mu_Y$ respectively; 
in our case $X$ is the number of lost logs and $Y$ is the total number of logs. 
Let $s_X, s_Y,$ and $\sigma_{X, Y}$ be the standard deviation of $X$, 
the standard deviation of~$Y$, and the covariance of $X$ and $Y$ respectively. 
Then the confidence interval of $X / Y$ can be (approximately) computed via
\[
\rho \pm \rho z \cdot  \sqrt{ \Bigl( \frac{s_X}{\mu_X} \Bigr)^2+ \Bigl( \frac{s_Y}{\mu_Y} \Bigr)^2 - \frac{2\sigma_{X, Y}}{\mu_X \cdot \mu_Y}}   \enspace, 
\]
where $\rho = \mu_X / \mu_Y$ is  the ratio of the two means, 
and $z$ is the Student $t$-value threshold with degree of freedom $n - 1$ for the desired confidence~level. 
As reported in Table~\ref{table:data_loss}, the 90\% confidence intervals in our case are consistently small across benchmarks, indicating reliable measurements and supporting the following key observations:

\begin{itemize}[leftmargin=*,noitemsep,topsep=0pt]
    \item \textbf{\Sys vs. \qltwo:} \Sys consistently outperforms \qltwo~\cite{281386} in data loss, with \qltwo exceeding 90\% loss in most benchmarks due to reliance on Auditd~\cite{linuxaudit}, which discards logs under load~\cite{Sekar2023eAA, nodrop}. When XMAC blocks logs in the kernel, Auditd cannot flush them to user space, leading to near-complete data loss in repeated high-load tests.

    \item \textbf{\Sys vs. \edm:} Simply integrating cryptographic protections into \ed, as done in \edm, results in high data loss and overhead. This shows that cryptographic features must be co-designed with system architecture. \Sys achieves this through tight integration with kernel memory management and efficient log handling.

    \item \textbf{\Sys vs. \ed:} While \ed uses weighted scheduling and a two-level cache to limit data loss, \Sys goes further by adding a two-level time controller, eBPF-aware padding, and compact encoding. This design maintains data integrity and low loss even under extreme load.
\end{itemize}
To evaluate resilience under peak load, we use only stress-test benchmarks for this RQ, as they offer controlled conditions to measure data loss during extreme system pressure. Real-world application benchmarks, though valuable for general performance, impose lighter workloads and are less effective at exposing data retention limits. At the same time, all results are reported with the thread count set to 16, as this represents the heaviest load. The trend is similar for all other settings.

\begin{table*}[!t]
    \centering
		\small
    \caption{Data loss comparison using \stresstest benchmarks.}
    \setlength{\tabcolsep}{2.5pt}
    \begin{tabular}{lccccccc}
    \toprule[1pt]
    \raisebox{-1.8ex}[0pt][0pt]{\textbf{\makecell{System}}} & \multicolumn{7}{c}{\textbf{Benchmarks (\%)}} \\
    \cmidrule(lr){2-8}
     & \textbf{Postmark} & \textbf{shbm} & \textbf{tar} & \textbf{find} & \textbf{httperf} & \textbf{rdwr} & \textbf{kernel} \\
    \midrule
    \qltwo & 93.92 $\pm$ 3.44 & 90.04 $\pm$ 2.18& 96.41 $\pm$ 1.79& 97.96 $\pm$ 1.54& 95.20 $\pm$ 3.61& 98.03 $\pm$ 1.05& 30.92 $\pm$ 2.35\\
    \edm & 35.67 $\pm$ 2.77& 29.58 $\pm$ 2.51& 40.17 $\pm$ 3.23& 48.90 $\pm$ 4.10& 34.64 $\pm$ 2.45& 52.11 $\pm$ 3.85& 9.75 $\pm$ 0.94\\
    \ed & 1.94 $\pm$ 0.18& 0.16 $\pm$ 0.03& 2.41 $\pm$ 0.18& 2.65 $\pm$ 0.26& 0.87 $\pm$ 0.07& 4.94 $\pm$ 0.46& 0.00 $\pm$ 0\\
    \rowcolor[gray]{.9} \Sys & 0.37 $\pm$ 0.03 & 0.14 $\pm$ 0.01 & 1.52 $\pm$ 0.12& 1.91 $\pm$ 0.17& 0.52 $\pm$ 0.06& 2.17 $\pm$ 0.12& 0.00 $\pm$ 0\\
    \bottomrule[1pt]
    \end{tabular}
    \label{table:data_loss}
\end{table*}

\subsection{RQ3: Optimization Breakdown}
\label{sub:opt_perf}

There are two major architecture optimizations in our design: (i) using per-core tags instead of a single tag, and (ii) using a two-level time controller in the log buffering. 
To quantify their contributions in improving performance, 
we use a full factorial design analysis.
 Below, we briefly describe the procedure of the analysis; 
one can see, for example,~\cite[Chapter 18]{Book:Analysis} for further details.

\heading{Full factorial design.} Here we have two factors in the design, each of two levels: 
signing (single tag or per-core tags) and buffering  (frequent versus occasional pushes). 
The \edm system,  for example, uses the single-tag signing and frequent pushes, 
whereas \Sys uses per-core tags and occasional pushes. 
To understand the effect of these two factors and their interaction, 
one would need to compare performance of all possible $2 \times 2 = 4$ systems.
We will use the Postmark benchmark where each experiment  has  $r = 30$ replications.  
This is known as a full $2^2 \times r$ factorial design. 
The benchmark has~5 setting for $1, 2, 4, 8, 16$ threads. 
Below, we describe how to analyze data for each setting.

Let $\Xs = -1$ if one signs with per-core tags, and $\Xs = 1$ otherwise. 
Let $\Xc = -1$ if one uses occasional-push caching, and $\Xc = 1$ otherwise. 
The running time of each system is modeled via a linear regression 
\( y = Q_0 + \Qs \Xs + \Qc \Xc + \Qi \Xs \Xc \enspace, \)
where $Q_0$ is the mean running time of the four systems,
$\Qs$ is the effect of signing, $\Qc$ is the effect of buffering, 
and $\Qi$ is the effect of the interaction of the two factors. 

The benchmarks will provide the average running time of each of the four systems, 
meaning that we have a system of four linear equations, 
 which is enough to determine the variables $Q_0, \Qs, \Qc, \Qi$. 
To determine the allocation of variation, 
we compare each actual running time $Y_k$ 
and the predicted time $y_k$ of the linear regression. 
Let 
\(
     \Se  = \sum_{k = 1}^{4r} (Y_k - y_k)^2
\)
be the total square error. 
The signing, buffering, and their interaction help explain a fraction $\Fs, \Fc, \Fi$ respectively in the variation, 
where for each $k \in \{s, b, i\}$, 
\[ \Fx = \frac{\Qx^2}{\Se/4r + \Qs^2 + \Qc^2 + \Qi^2} \enspace. 
\] 
The remaining (unexplained) fraction $F_e = 1 - (\Fs + \Fc + \Fi)$ of variation is due to  errors in the experiments.

\subsubsection{Runtime Overhead}

The results are reported in Table~\ref{table:reg}.
The unexplained fraction $F_e$ is negligible, suggesting a strong model fitness.
The two factors have essentially no interaction because signing improves CPU operations,
whereas buffering improves I/O ones.
Signing is the dominant factor; for example it explains nearly 80\% of the variation
for the setting of 16 threads.
As expected, the relative contribution of signing increases when the level of concurrency increases.
Notably, the value of \( F_i \) remains averagely 1.75\%, suggesting that the combination of the two factors has a positive effect, but not a significantly large one. 
Even if there is only one thread, \( F_s \) still plays a dominant role, because having a single thread does not necessarily mean that only one CPU core is used. 
Although only one core is responsible for running the process at any given moment, it may switch between cores based on system scheduling. In per-core tag versus single-tag systems, the data structures for storing system secrets differ fundamentally: one utilizes a Per-CPU Array, while the other relies on a global structure in kernel space (e.g., \texttt{BPF\_MAP}). Even with thread=1, this structural difference leads to performance variation, making \( F_s \) still dominant under this setting.

\begin{table}[!t]
  \centering
  \footnotesize
  \caption{Allocation of variation on running time. Each reported number is a percentage.
  The unexplained fraction $F_e$ is negligible in all settings.}
  \resizebox{0.45\textwidth}{!}{\begin{tabular}{l r r r r r}
  \toprule
  &  \textbf{1 thread} & \textbf{2 threads} & \textbf{4 threads} & \textbf{8 threads} & \textbf{16 threads} \\
  \midrule
  $\Fs$ & 52.49 & 56.84 & 61.75 & 72.17 & 80.15 \\
  $\Fc$ & 43.92 & 41.13 & 36.62 & 26.92 & 19.24 \\
  $\Fi$ & 3.59 & 2.03 & 1.63 & 0.91 & 0.61 \\
  \bottomrule
  \end{tabular}
  }
  \label{table:reg}
  \vspace{-2ex}
  \end{table}

  \subsubsection{Data Loss}
  \label{sub:opt_dataloss}
  \begin{table}[!t]
  \centering
  \footnotesize
  \caption{Allocation of variation on the number of lost logs. Each reported number is a percentage. 
  The unexplained fraction $F_e$ is negligible in all settings.}
  \resizebox{0.45\textwidth}{!}{\begin{tabular}{l c c c c c}
  \toprule
   & \textbf{1 thread} & \textbf{2 threads} & \textbf{4 threads} & \textbf{8 threads} & \textbf{16 threads} \\
  \midrule
  $\Fs$ &26.94  &29.85  &31.24  &34.95  &42.92  \\
  $\Fc$ &72.53 &69.10  & 66.31 &62.28  &53.45  \\
  $\Fi$ &0.53  &1.05  &2.45  &2.77  &3.63  \\
  \bottomrule
  \end{tabular}
  }
  \label{table:regd}
  \vspace{-2ex}
  \end{table}

Here we give a breakdown of the contributions of the two optimization factors (signing and buffering) on data loss via a full factorial design
with the Postmark benchmark.  
However, one can't directly do a linear regression on the data loss ratio, because this involves adding ratios of different denominators, 
which is meaningless.  
Instead, we do a linear regression on the number of lost logs. At the same time, although fast push clearly increases overhead, it is beneficial for reducing data loss, making it unnecessary to analyze the impact of fast push versus occasional push on data loss in this context. The key factor we designed to address data loss under high loads, the two-level time controller, essentially serves as an accelerator for cache pushes. Thus, we further refine the definition of \( x_b \): Let \( x_b = -1 \) if using occasional-push caching with a two-level time controller, and \( x_b = 1 \) if using standard occasional-push caching.

The results are given in Table~\ref{table:regd}. 
Again, the interaction is small. It is evident that the two-level time controller plays a major role in reducing data loss, with its contribution showing a consistent pattern as the thread count changes. When the number of threads increases, the influence of per-core tags also grows; however, the two-level time controller remains dominant, accounting for nearly 71\% when the thread count is only 1. Notably, as the load continues to increase, the combined effect of per-core tags and the two-level time controller also becomes more pronounced, indicating that their combined impact is more effective under high-load conditions.

\subsection{RQ4: Overhead \& Data Loss using \SysR}
\label{sub:cmp_sysr}

\begin{figure}[t!]
    \centering
    \subfigure[PostMark]{
        \centering
        \includegraphics[width=0.22\textwidth]{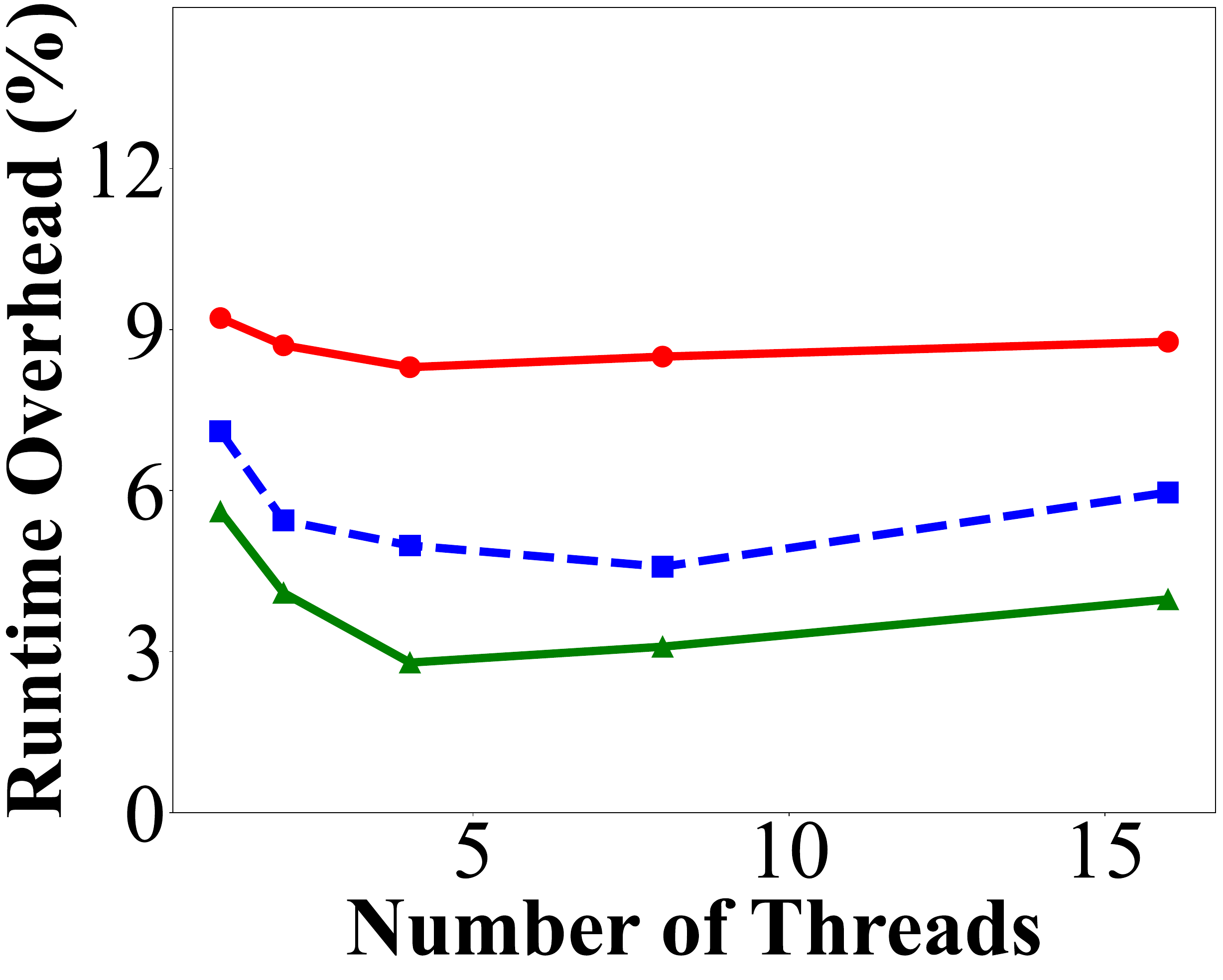}
    }
     \subfigure[shbm]{
        \centering
        \includegraphics[width=0.22\textwidth]{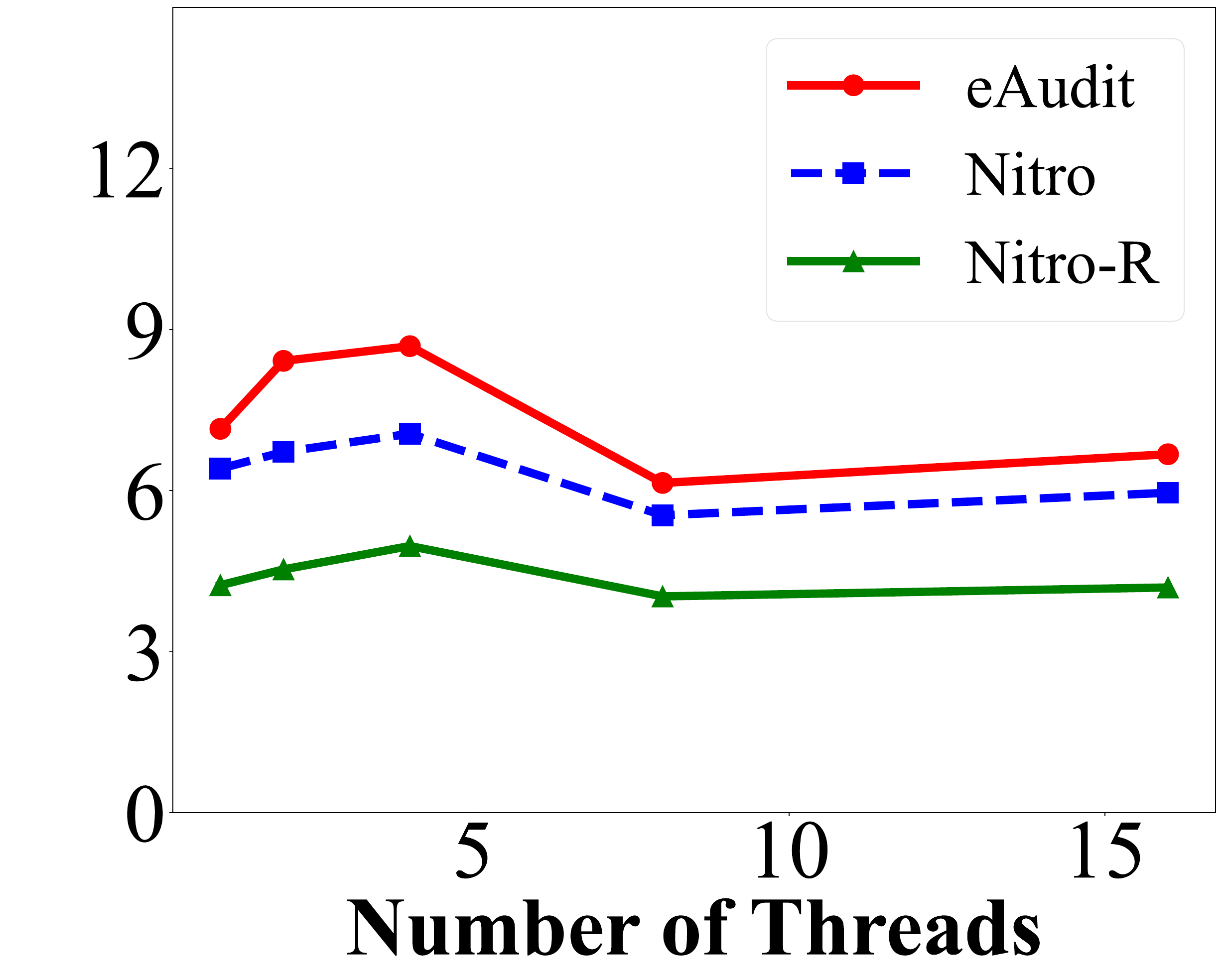}
    }
    \caption{Comparison of \SysR, \Sys, and eAudit on two benchmarks. The remaining results are provided in Appendix~\ref{D:sr_cmp}.} 
    \label{fig:sysr_cmp_all}
    \vspace{-3ex}
\end{figure}

To thoroughly evaluate \SysR, we conducted comparative tests on all \stresstest and real-world benchmarks, using \Sys and eAudit~\cite{Sekar2023eAA} as baselines. Due to space constraints, we present results for \texttt{shbm} and \texttt{PostMark}, with the remaining results included in Appendix~\ref{D:sr_cmp}. The additional results exhibit trends consistent with those shown here. Figure~\ref{fig:sysr_cmp_all} shows that \SysR achieves an average runtime overhead reduction of 24\% on \stresstest benchmarks and 11\% on real-world benchmarks, relative to \Sys, which already maintains low overhead. It also reduces redundant logs by 87.91\% on \stresstest workloads and 49.76\% on real-world ones. With these improvements, \SysR surpasses the performance of eAudit~\cite{Sekar2023eAA}. Its use of an in-kernel LRU hash table enables efficient log reduction with minimal overhead, maintaining high performance under load. By removing duplicate logs before they reach user space, \SysR reduces cache contention and pressure on downstream components. This results in lower runtime overhead and improved resilience to data loss, without affecting data integrity. Under an expanded definition of data loss that considers the proportion of effective log data lost, \SysR is the only system to achieve zero loss across all benchmarks. These results were consistent across repeated trials and are summarized in Table~\ref{table:data_loss}.
 
\section{Discussion and Future Directions}
\label{sec:discussion}

\noskipheading{Combining \qltwo and \ed.} A natural question that arises is whether the security benefits of QuickLog2 can be combined with the performance advantages of eAudit. However, our experiments (Section~\ref{s:eval}) show that such integration is practically prohibitive due to conflicting design principles: 1) eAudit uses small buffers and frequent log pushes to reduce the tamper window, leading to high I/O overhead, while QuickLog2 relies on large buffers and batch processing for cryptographic efficiency; combining them forces frequent signing, negating batching benefits. 2) QuickLog2’s AES-based MAC depends on AES-NI hardware acceleration, which is incompatible with eBPF. 3) eAudit supports multi-core logging, but QuickLog2 performs sequential cryptographic processing, creating bottlenecks when combined.

\heading{Cross-Platform Portability.} Although designed for Linux, \Sys could be adapted for Windows, as the Windows platform now supports eBPF~\cite{microsoft2024ebpf}. At the same time, Nitro's core design, including Per-CPU Array and ring buffer, is also supported by eBPF for Windows~\cite{ebpf_windows_file_reference,ebpf_windows_official_intro}. Therefore, porting Nitro to other systems does not require major design changes. However, due to fundamental differences between Windows and Linux, the migrated Nitro requires modifications to its event capture hooks, such as the names of entry events~\cite{ebpf_windows_tutorial}. The closed-source nature of Windows limits low-level kernel event monitoring, with most monitoring capabilities focusing on network traffic. Users can integrate Event Tracing for Windows (ETW) to assist in analyzing the runtime behavior of eBPF programs on Windows~\cite{ebpf_windows_ETW}. Secondly, the log encoding scheme also requires adaptation. Since Nitro’s padding strategy relies on analyzing parameters of the monitored targets, the encoding rules must be redefined for the new platform. Specifically, known-length parameters need to be combined to meet the input length requirement of XLog’s MAC, which represents an engineering effort. Lastly, the memory usage constraints differ across platforms. With the recent update to eBPF for Windows, which relaxes the size limitations~\cite{microsoft2024ebpf}, the optimal configuration proposed in this paper is now supported.
 While outside the focus of this paper, adapting \Sys for Windows environments is a practical future direction that could broaden its applicability in diverse security contexts.

\heading{Choosing between \Sys and \SysR.} End users can choose \Sys or \SysR based on application needs. \SysR performs log reduction in the kernel. Compared to \Sys, it has lower runtime overhead and less data loss. It also removes redundant logs, such as repeated file reads, reducing storage use. \SysR fits scenarios with lower log sensitivity and stronger performance or storage demands, such as large-scale cloud services, CDNs, or internal testing. For use cases needing higher log integrity, such as enterprise systems, financial platforms, government facilities, or critical infrastructure, \Sys is preferred.

\heading{Limitations.} While \Sys and \SysR offer efficient, tamper-evident logging, some limitations remain. First, both systems support auditing and post-incident investigation, not runtime attack prevention, which aligns with other forensic tools~\cite{kennylog,Sekar2023eAA,281386,linuxaudit}. Second, \SysR boosts performance and cuts storage by filtering high-frequency, low-value events. This may reduce temporal precision, but prior work~\cite{auditrim} shows limited impact on forensic usefulness. Third, \Sys depends on stable BPF tracepoints and CO-RE for cross-version compatibility. This lowers maintenance in most cases, but major kernel changes may still require BPF updates~\cite{zhong2025revealing}.
 \section{Related Work}
\label{s:relwk}
\heading{Hardware-Assisted secure logging.} While software-only solutions like \Sys can detect attacks, they cannot prevent an adversary from wiping log data from disk to erase evidence. A complementary approach involves using special-purpose hardware for secure log storage. For instance, \hardlog~\cite{hardlog} uses a RockPro64 development board to synchronously record predefined critical system calls. \omini~\cite{omini} employs an NXP IMX8MQ-EVK board for more comprehensive system call logging and ensures synchronous log availability. \sgx~\cite{kbl+2017} leverages Intel SGX to protect code and data in memory and on disk from tampering, even by privileged software. \custos~\cite{custos} also uses Intel SGX, integrating cryptographic mechanisms to further enhance logging security. More recently, HitchHiker~\cite{hitchhiker} leverages EL3-based memory permission switching on ARM platforms to improve log protection assurance. While it significantly reduces the TCB and protection delay, its reliance on periodic permission switching introduces non-negligible runtime overhead under log-intensive workloads.

\heading{eBPF-based logging system.} The ease of deployment offered by eBPF has made it increasingly popular for building modern logging systems. Notable examples include tracee~\cite{tracee2}, Tetragon~\cite{Tetragon}, Sysdig~\cite{Sysdig}, \ed~\cite{Sekar2023eAA}, and Falco~\cite{Falco}, all of which are either built from scratch or significantly enhanced using eBPF. However, none of these systems is tamper-evident or fully eliminates the window for log tampering. With \Sys, we introduce the first tamper-evident logging system based on eBPF, enhancing the security guarantees of prior systems without sacrificing performance.

\heading{Log Reduction Schemes.} Numerous log reduction schemes, such as those in~\cite{inam2022sok,Xu:2016,elise,tang2018nodemerge,hossain+depend,auditrim,swift2020}, focus on minimizing audit log volume by offloading processing to user space, which introduces transfer overhead. In contrast, \Sys leverages in-kernel log reduction to eliminate the overhead of log transfers. This paper specifically implements Auditrim's~\cite{auditrim} log reduction strategy within the kernel to minimize performance costs, achieving a log reduction ratio similar to AudiTrim's. Future adaptations of other log reduction schemes into the kernel could further enhance \Sys's efficiency in managing log data with reduced overhead.
 \section{Conclusion}
\label{s:conclusion}

This paper presents \Sys, the first tamper-evident logging system fully implemented in eBPF. By co-designing the cryptographic and systems architecture, \Sys addresses longstanding challenges in logging performance, granularity, and maintainability. Our enhanced variant, \SysR, further reduces overhead by incorporating in-kernel log reduction, demonstrating that practical, high-integrity logging is achievable even under high system loads.
 
\section{Acknowledgment}
\label{s:ack}

We sincerely thank the anonymous shepherd and reviewers for their insightful feedback on this work. This material is based upon work supported by the National Science Foundation (NSF) under grant numbers CNS-2339483 and CNS-2046540, and by the Commonwealth Cyber Initiative (CCI) Award. 

\balance
\bibliographystyle{abbrvnat}
{\footnotesize\setlength{\bibsep}{3pt}
\bibliography{arxiv.bbl}}

\begin{appendix}
    
\section{Proof of \propref{P:agg}} \label{A:agg_security_proof} 

Consider the following games $P_1$ and $P_2$. 
Game $P_1$ corresponds to game  $\Game^{\forge}_{\AMAC}(\advA)$. 
Game $P_2$ is similar to game $P_1$, but each call to $G(K_i, \cdot)$ is replaced
by the corresponding call to a truly random function $f_i: \bits^* \to \bits^\tau$. 
 Note that in game $P_2$, the correct aggregate tag $T^*_r$ of the forgery messages is uniformly random over $\bits^\tau$, independent of whatever the adversary~$\advA$ receives.
    Thus the chance that it can guess $T^*_r$ correctly is $2^{-\tau}$. In other words, $\Pr[P_2(\advA)] = 2^{-\tau}$.

To bound the gap between the two games above, consider the following adversary $\advB$ attacking the (multi-user) PRF security of $G$. 
It runs $\advA$ and simulates game $P_1$. 
However, for each call to $G(K_i, \cdot)$, adversary $\advB$ instead makes the corresponding oracle query $\FnO(i, \cdot)$. 
Hence game $P_1(\advA)$ corresponds to  game $\Game^{\muprf}_G(\advB)$ with challenge bit $1$, 
whereas game $P_2(\advA)$ corresponds to game $\Game^{\muprf}_G(\advB)$ with challenge bit $0$. 
Hence 
\[ 
\Pr[P_1(\advA)] - \Pr[P_2(\advA)] = \Pr[\Game^{\muprf}_G(\advB)] = \Adv^{\muprf}_G(\advB) \enspace. 
\] 
Hence, 
\begin{eqnarray*}
\Adv^{\forge}_\AMAC(\advA)  &=& \Pr[P_1(\advA)]  \\
&=& \Pr[P_1(\advA)] - \Pr[P_2(\advA)] + \Pr[P_2(\advA)] \\
&=& \Adv^{\muprf}_G(\advB)  + \Pr[P_2(\advA)] \\
&\leq& \Adv^{\muprf}_G(\advB) + 2^{-\tau} \enspace. 
\end{eqnarray*}

\section{Proof of \thref{T:XLog_security}} \label{A:XLog_security_proof}
    
       Without loss of generality, assume that $\advA$ alters the logs. 
    Since the adversary $\advA$ receives the state $S_{q}$ and key $K_{q}$, it can compute any  key $K_i$ (with $i > q$).
    Thus without loss of generality, we assume that $\advA$ only generates at most $q$ forgery messages.
    Indeed, suppose that~$\advA$ outputs $(M'_1, \ldots, M'_r, T')$ with $r > q$.
    Then it could  instead produce $(M'_1, \ldots, M'_q, T'_q)$ to win with the same advantage,
		where~$T'_q$ is computed as follows. Let $T'_r = T'$, and for each $i$ from~$r$ down to $q + 1$, let $T'_{i - 1} = T'_i \combinv G(K_{i - 1}, M'_i)$. 
		Hence from now on, we assume that $r \leq q$.

		 \begin{figure}[t!]
    \twoColsNoDivide{0.43}{0.2}
    {
    \underline{\textbf{Adversary} $\advD()$} \\[2pt]
       $(M_1, \ldots, M_q, \sigma) \getsr \advA$ \\
			 Return $(M_1, \ldots, M_q, \sigma)$ \\[4pt]
		 \underline{\textbf{Adversary} $\advD(\sigma, T)$} \\[2pt]
       $X_1, \ldots, X_q \getsr \bits^n$;~ $S, K \getsr \bits^n$ \\
    $(M'_1, \ldots, M'_r, T') \getsr \advA(S, K, X_1, \ldots, X_q, T, \sigma)$ \\
    Return $(M'_1, \ldots, M'_r, T')$
    }
    {
    }
	   \vspace{-1ex}
		 \caption{Constructed adversary~$\advD$ in the proof of \thref{T:XLog_security}.}
    \vspace{-1ex}
			 \label{fig:advD}
      \end{figure}

		Consider the following games $G_1$ and $G_2$. 
		Game $G_1$ corresponds to game $\Game_{\XLog[\AMAC, F]}^{\faa}(\advA)$. 
		Game $G_2$ is similar to game $G_1$, but all the keys and states and tag masks are sampled uniformly at random.
		To bound the gap between the two games, 
		consider the following adversary $\advB$ attacking the (multi-user) PRF security of $F$. 
It runs~$\advA$ and simulates game $G_1$. 
However, for each call to $F(S_{i - 1}, \cdot)$, adversary $\advB$ instead makes the corresponding oracle query $\FnO(i, \cdot)$. 
Hence game $G_1(\advA)$ corresponds to  game $\Game^{\muprf}_F(\advB)$ with challenge bit $1$, 
whereas game $G_2(\advA)$ corresponds to game $\Game^{\muprf}_F(\advB)$ with challenge bit $0$. 
Hence 
\[ 
\Pr[G_1(\advA)] - \Pr[G_2(\advA)] = \Pr[\Game^{\muprf}_F(\advB)] = \Adv^{\muprf}_F(\advB) \enspace. 
\] 
To bound $\Pr[G_{2}(\advA)]$, consider the adversary $\advD$ attacking $\AMAC$ as described in \figref{fig:advD}. 			
		It perfectly simulates game $G_2(\advA)$, and thus
		\[ 
		\Adv^{\forge}_{\AMAC}(\advD) = \Pr[G_2(\advA)] \enspace. 
		\] 
		Summing up, 
		\begin{eqnarray*}
		\Adv^{\faa}_{\XLog[\AMAC, F]}(\advA) &=& \Pr[G_1(\advA)] \\
		&=& \Pr[G_1(\advA)]- \Pr[G_2(\advA)] + \Pr[G_2(\advA)] \\
		&=& \Adv^{\muprf}_F(\advB) + \Pr[G_2(\advA)] \\
		&=& \Adv^{\muprf}_F(\advB) + \Adv^{\forge}_{\AMAC}(\advD) \enspace. 
		\end{eqnarray*}

\section{Description of Benchmarks}
\label{app:benchmarks}

We evaluate our system using two categories of workloads: \textit{Stress-test Benchmarks} and \textit{Real-world Benchmarks}.
The \textit{Stress-test Benchmarks} in Table~\ref{tab:micro} are adapted from prior work~\cite{Sekar2023eAA} and are designed to evaluate system behavior under high-load conditions. These benchmarks target specific subsystems such as file system metadata operations (e.g., \texttt{postmark}, \texttt{find}), process management (\texttt{shbm}), system call throughput (\texttt{rdwr}), network stack performance (\texttt{httperf}), and large-scale build workloads (\texttt{kernel}). Together, they provide a controlled environment to stress the system and measure overhead, data loss, and scalability under pressure.
The \textit{Real-world Benchmarks.} in Table~\ref{tab:real} capture common workloads in modern systems, including web browsing, web serving, caching, cryptographic operations, file compression, and file system activity. They reflect practical deployment scenarios for logging and auditing systems, and help evaluate the system’s effectiveness and overhead under realistic conditions.

\begin{table}[H]
    \small
\centering
\caption{Stress-test benchmarks in this paper.}
\label{tab:micro}
\renewcommand{\arraystretch}{1.2}
\begin{tabularx}{0.46\textwidth}{>{\raggedright\arraybackslash}p{0.11\textwidth}X}
   \toprule[1pt]
\textbf{Name} & \textbf{Description} \\ 
\midrule
Postmark~\cite{postmark} & Simulation of a mail server. \\
tar~\cite{Sekar2023eAA} &  Using \texttt{tar} to archive /usr/lib. \\
shbm~\cite{Sekar2023eAA} & A shell script executing \texttt{echo} repeatedly. \\ 
find~\cite{Sekar2023eAA} & Using \texttt{find} to print file names in /usr	. \\ 
httperf~\cite{Sekar2023eAA} &  A benchmark for web servers. \\ 
rdwr~\cite{Sekar2023eAA}  & \parbox[t]{\linewidth}{A C-program calling \texttt{read} and \texttt{write}.} \\ 
kernel~\cite{Sekar2023eAA} & \parbox[t]{\linewidth}{Compiling the Linux kernel.} \\
  \bottomrule[1pt]
\end{tabularx}
\end{table}

\begin{table}[H]
\centering
\small
\caption{Real-world benchmarks in this paper.}
\label{tab:real}
\renewcommand{\arraystretch}{1.2}
\begin{tabularx}{0.46\textwidth}{>{\raggedright\arraybackslash}p{0.11\textwidth}X}
  \toprule[1pt]
\textbf{Application} & \textbf{Benchmarks} \\ \midrule
Firefox~\cite{firefox} & Speedometer benchmark for web browser~\cite{firebench}. \\ 
Nginx~\cite{nginx} &  Benchmark for a Nginx web server~\cite{abbench} (with 10,000 requests for a 1KB file
under 12 concurent threads in a local environment). \\ 
Redis~\cite{redis} & Benchmark for a Redis server~\cite{redisbench}. \\ 
OpenSSL~\cite{openssl} & Benchmark for cryptographic operations~\cite{phoronix}. \\ 
7zip~\cite{7zip} &  Benchmark for file compression~\cite{phoronix}. \\ 
lmbench & Benchmark for file creation and deletion (0K and 10K) in the lmbench benchmark suite~\cite{lmbench}.  \\
\bottomrule[1pt]
\end{tabularx}
\end{table}

\section{List of Syscalls to Monitor}
\label{app:syscalls}

Below is the list of syscalls that we monitor:

\begin{itemize}
\item \emph{Privilege escalation and tampering:} execve, execveat,  setresuid, setuid,
setfsuid, setfsgid, setgid, setregid, \mbox{setresgid}, kill, tgkill, tkill, ptrace.

\item \emph{Process provenance:} clone, clone3, exit, exit\_group,  mmap, mprotect, vfork.

\item \emph{File name and attribute change:} unlink, unlinkat, chmod, fchmod, fchmodat, chdir, fchdir.

\item \emph{Data endpoint creation:} connect, accept, accept4, open, openat, creat, socket.

\item \emph{Sends and receives:} sendto, recvfrom, sendmmsg, sendmsg, recvmsg, recvmmsg.

\item \emph{File descriptor operations:}  dup, dup2, dup3, pipe, pipe2, tee, rmdir,   truncate, ftruncate, mkdir, mkdirat, mknod, mknodat, socketpair.

\item \emph{Reads and writes:} read, readv, pread64, preadv, preadv2,  write, writev, pwrite64, pwritev, pwritev2,   splice,   vmsplice.

\item \emph{Others:} getpeername.
\end{itemize}

\section{Detailed Algorithm of \SysR}

\SysR implements a verifier-safe, in-kernel log reduction algorithm using eBPF to eliminate redundant system log entries in real time. Algorithm~\ref{alg:SysR} presents the detailed workflow. For each incoming log entry $e$ from the raw log stream $R_L$, \SysR extracts a structured semantic key $\alpha$, which consists of the process ID, system call name, and its arguments. This key represents the semantic identity of the event.

To detect duplicates efficiently, \SysR maintains an in-kernel Least-Recently-Used (LRU) hash table $H$, which maps each key $\alpha$ to the timestamp of its most recent occurrence. If the same key appears again within a defined time window $T_W$, the event is considered redundant and dropped. Otherwise, the current timestamp is stored or updated in $H$, and the entry is added to the reduced log $R_D$.

The algorithm ensures real-time performance and verifier compatibility by avoiding dynamic memory allocation and unbounded loops. Instead, it uses fixed-size structures, bounded iteration, and simple branching logic. Additionally, the LRU eviction policy of $H$ prevents memory overflow, enabling scalable, long-term operation without kernel modification. The log reduction window $T_W$ is configurable and can be dynamically adapted to system conditions if needed, although a static value (e.g., 1 second) is used in the default configuration. This design ensures that only unique entries are retained and passed forward for cryptographic processing, minimizing both computational and I/O overhead without sacrificing log integrity or completeness.

\begin{algorithm}[h]
    \SetAlgoNoEnd
    \small
    \DontPrintSemicolon
    \SetKwInOut{Input}{Input}
    \SetKwInOut{Output}{Output}
    
    \Input{Raw Log: $R_L$, Time Window: $T_W$}
    \Output{Reduced Log: $R_D$}

    $H \leftarrow \text{LRU Hash Table}$ \tcp{Initialize or reload LRU Hash Table $H$}
    
    \ForEach{$e$ in $R_L$}{
      \tcc{Map log entry to a structured format}
      $\alpha \leftarrow  (\text{PID}, \text{Name}, \text{Args}) \leftarrow e$\;
      
      \BlankLine
      \tcc{Check if entry is a duplicate within the specified time window}
      \If{$\alpha \in H$}{
          \If{$(\text{Time}_{\text{current}} - H[\alpha]) \leq T_G$}{
            
            Drop $e$ \tcp{Duplicate entry found, drop it}
            Continue\;
          }
          \Else{
            $T_c \leftarrow$ getCurrentTime()\;
            Update $(\alpha: T_c)$ in $H$\;
          }
      }
      \Else{
        $T_c \leftarrow$ getCurrentTime()\;
        Insert $(\alpha: T_c)$ into $H$\;
      }

      \BlankLine
      
      $R_D \leftarrow R_D \cup \text{process}(e)$ \tcp{Add unique entry to reduced log}
    }
    
    \Return{$R_D$}\;
    
    \BlankLine
    \caption{In-Kernel Log Reduction}
    \label{alg:SysR}
\end{algorithm}

\section{RQ5: Memory Usage}
\label{sec:memory_usage}

\begin{figure}[h]
	\centering
		\includegraphics[width=0.35\textwidth]{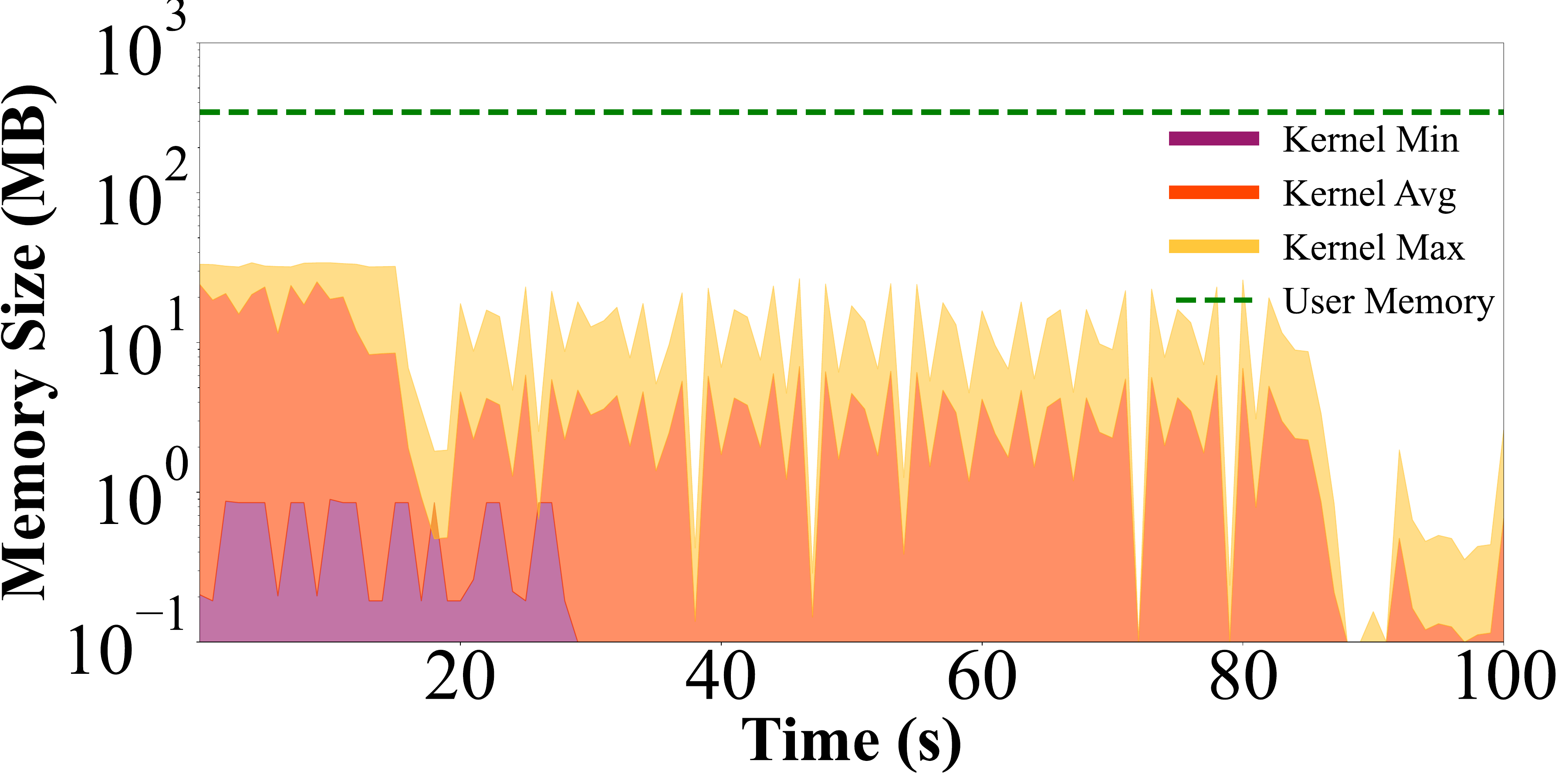}
	\caption{Kernel and user space memory usage of \Sys across all \stresstest benchmarks. Kernel memory tracks in-kernel data structures; user memory is \Sys's Resident Set Size (RSS).}
	\label{fig:sys_memory}
  \vspace{-2ex}
\end{figure}

We analyze the memory consumption of \Sys from two perspectives: kernel space and user space, as shown in Figure~\ref{fig:sys_memory}. In the kernel, memory allocation and actual memory usage are distinct concepts. \Sys relies on eBPF-compliant data structures, and their sizes must be specified at compile time. As a result, regardless of how much memory is actually used at runtime, these structures are preallocated and persist in memory throughout execution. Specifically, \Sys allocates a Per-CPU Array of size $S_p$ for each logical core and a shared ring buffer of size $S_r$. Given our configuration with $N = 36$, $S_p = 32$ KB, and $S_r = 64$ MB, the total allocated kernel memory can be computed as $N \times S_p + S_r$, which amounts to approximately 65 MB. This value represents the upper bound on memory that can be consumed, but not the actual memory usage at runtime.
In Figure~\ref{fig:sys_memory}, the actual kernel memory usage of \Sys increases sharply after the \stresstest benchmarks begin. It then fluctuates for a period before gradually decreasing to a lower level and stabilizing. In user space, we track the Resident Set Size (RSS) to reflect the memory usage of \Sys. \Sys relies on the Python runtime and BCC bindings to receive and process logs~\cite{ebpf-bcc}. This component is initialized once and maintains stable memory usage throughout execution. Our evaluation on \stresstest benchmarks shows that the user space memory consumption for \Sys consistently remains around 345MB, regardless of workload intensity or logging frequency.

\section{RQ6: Comparison with NoDrop} \label{sec:nodrop-cmp}

\begin{figure*}[t!]
    \centering
    \includegraphics[width=0.85\textwidth]{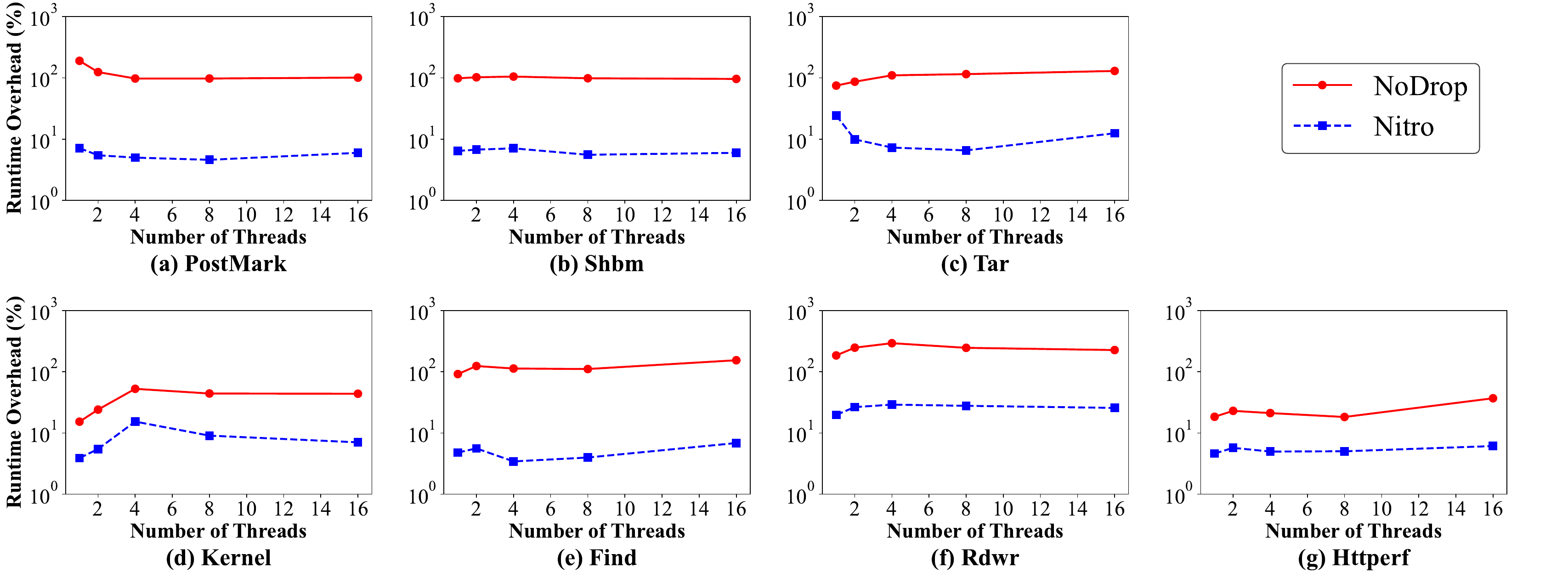}
    \caption{Runtime overhead of \nodrop and \Sys.}\label{fig:nodrop_cmp_all}
    \vspace{-2ex}
\end{figure*}

To evaluate the efficiency of \Sys against \nodrop~\cite{nodrop}, a state-of-the-art tamper-proof logging system, we ran all seven stress-test benchmarks on an Intel-based machine, as \nodrop requires Intel MPK support. Other system configurations were kept identical to our main evaluation setup.

Figure~\ref{fig:nodrop_cmp_all} presents the complete comparison results across all benchmarks. On average, \Sys incurs over 10× lower runtime overhead than \nodrop across all benchmarks. In I/O-heavy scenarios like \texttt{rdwr} and \texttt{find}, the overhead reduction reaches up to 25×.
This performance gap stems from key architectural differences. \nodrop uses Intel MPK for kernel-user isolation and relies on synchronized logging, resulting in frequent control transfers and coordination overhead. In contrast, \Sys employs a fully asynchronous, eBPF-based design, leveraging per-core logging and decoupled flushing to minimize syscall blocking and user space transitions.

While \nodrop guarantees zero data loss, we do not conduct a data loss comparison here, as that is its primary design goal. However, \Sys achieves near-zero loss (see Section~\ref{s:eval}) while offering significantly better performance. This tradeoff makes \Sys a more practical choice for high-throughput production environments.

\section{RQ7: Log tampering detection against real-world attacks} \label{app:real-world-attacks}

Theoretically, \secref{s:tma} shows that it's impossible to alter logs in \Sys without being detected, unless one can break the PRF security of Chaskey. 
Still,  it would be interesting to evaluate this on prior attacks. 
(This evaluation was in fact requested by reviewers of a prior submission to IEEE Security \& Privacy 2025.)
To showcase \Syss ability on detecting log tampering, we consider attack scenarios from~\cite{kennylog,custos}. 
All experiments are conducted in a controlled virtualized environment running Ubuntu 22.04 (with kernel version 6.5.0) and repeated 20 times.
We generate encrypted tags for every log to test \Syss ability to extract unmodified logs.

\heading{The attacks.} The first attack is based on the race attack in~\cite{kennylog} that led to CVE-2017-16995~\cite{cve_16995}. 
The original attacks generates 25 forensic-relevant syscalls but contains hardcoded memory offsets that cause errors on modern Ubuntu kernels. We modify it to fix this issue, ensuring that all syscalls execute in order.
To simulate the truncation attack, in each run, we sample a number $V \getsr \{1, \ldots, 25\}$ and only use the logs of the first $V$ syscalls. 

For the second attack, we adopt   the forensic tampering scenario from Custos~\cite{custos}, using the dataset from its attack evaluation section. 
This dataset originates from the DARPA TC dataset~\cite{Darpa_E3} and is labeled as ``Firefox Backdoor w/ Drakon In-Memory''. It consists of sequential log events mapping to specific kernel syscalls, including file writes, process execution, and network communications. 

\heading{Evaluation results.} In every case, \Sys  points out that the logs are altered, and correctly extracts the list of unmodified logs.

\section{RQ8: Log Correctness Analysis} \label{app:log-correctness}

Following prior work~\cite{protracer,inam2022sok,omegalog2020}, we validate the correctness of logs produced by \Sys by checking whether they preserve accurate causal relationships. Our goal is to confirm that asynchronous per-CPU logging does not compromise provenance integrity, and that the combination of timestamps and CPU core indexes suffices to reconstruct the original partial order of events. We adopt a methodology that combines online log collection with offline causality reconstruction, introduce formally defined reference baselines for comparison, and provide theoretical justification, as described below.

\heading{Baselines.}
We define one baseline: \( B_1 \), corresponding to a centralized logging configuration for \Sys. The baseline replaces the per-core asynchronous buffer with a centralized \texttt{perf\_buffer} that synchronously records all events in a globally ordered stream. All other aspects of the system remain unchanged. Let \( \mathcal{L}_{B_1} \) denote the logs generated under this baseline, and let \( \mathcal{L}_{\Sys} \) denote the logs produced by the default \Sys configuration with asynchronous per-CPU logging. This centralized baseline preserves a total order of system events and serves as ground truth for verifying the correctness of event relationships reconstructed from \( \mathcal{L}_{\Sys} \).

\heading{Online: Log Collection.}
The baseline configuration (\( B_1 \)) logs all events into a single, centrally ordered stream, ensuring global temporal alignment. In contrast, \Sys buffers events independently on each CPU core and flushes them asynchronously to user space.

\heading{Offline: Causal Relationship Reconstruction.}
We collect logs from both configurations while running the full suite of \stresstest benchmarks (Section~\ref{s:eval}), which naturally exercise a wide range of causal relationships, including process creation, inter-thread communication, I/O callbacks, and signal delivery.
From these logs, we analyze whether the expected causal relationships are preserved. In \( \mathcal{L}_{B_1} \), relationships are derived directly from the globally ordered stream. For \( \mathcal{L}_{\Sys} \), we merge per-core logs using timestamps and CPU core indexes to reconstruct the same set of causal dependencies. Specifically, we evaluate:
\begin{itemize}[leftmargin=*]
    \item Whether the same causal dependencies appear in \( \mathcal{L}_{\Sys} \) as in \( \mathcal{L}_{B_1} \);
    \item Whether any spurious, missing, or misordered relationships are introduced;
    \item Whether asynchronous logging leads to inconsistencies under concurrency.
\end{itemize}

\heading{Theoretical Justification.}
Let \( e_1 \) and \( e_2 \) denote two system events such that \( e_1 \prec e_2 \) in the baseline log \( \mathcal{L}_{B_1} \). Each event corresponds to a syscall record captured by the logger, and includes metadata such as the syscall name, related parameters, timestamp \( t_i \), and CPU core index \( c_i \).

Assuming (1) clocks are synchronized across cores (or clock skew is bounded and known), and (2) intra-core event order is preserved, we claim that the ordering of events in the reconstructed logs from \Sys, denoted as \( \mathcal{L}_{\Sys} \), can be recovered by lexicographic sort over \((t, c)\):

\[
    e_1 \prec e_2 \Rightarrow (t_1 < t_2) \lor (t_1 = t_2 \land c_1 < c_2)
\]

That is, if \( e_1 \) occurred before \( e_2 \) in the baseline, then the lexicographically ordered logs from \Sys will reflect the same precedence. Note that the secondary comparison on CPU core index does not imply a causal dependency, but rather serves as a deterministic tie-breaker to resolve simultaneous events with identical timestamps. In practice, such ties typically arise from concurrent, independent activities on different cores.

Across all \stresstest benchmarks, the reconstructed event relationships from \( \mathcal{L}_{\Sys} \) are consistent with those from \( \mathcal{L}_{B_1} \). All tested interaction patterns preserve their expected ordering. No violations were observed, such as reordered, missing, or spurious dependencies. Additionally, we performed similar experiments with \SysR and we observed consistent results. These results confirm that timestamps and CPU core indexes provide sufficient information to reconstruct correct event relationships, even under asynchronous per-core logging. By validating against well-defined baselines, we demonstrate that \Sys and \SysR maintain log correctness and provenance fidelity across diverse workloads.

\begin{figure}[t!]
    \centering
    \includegraphics[width=0.40\textwidth]{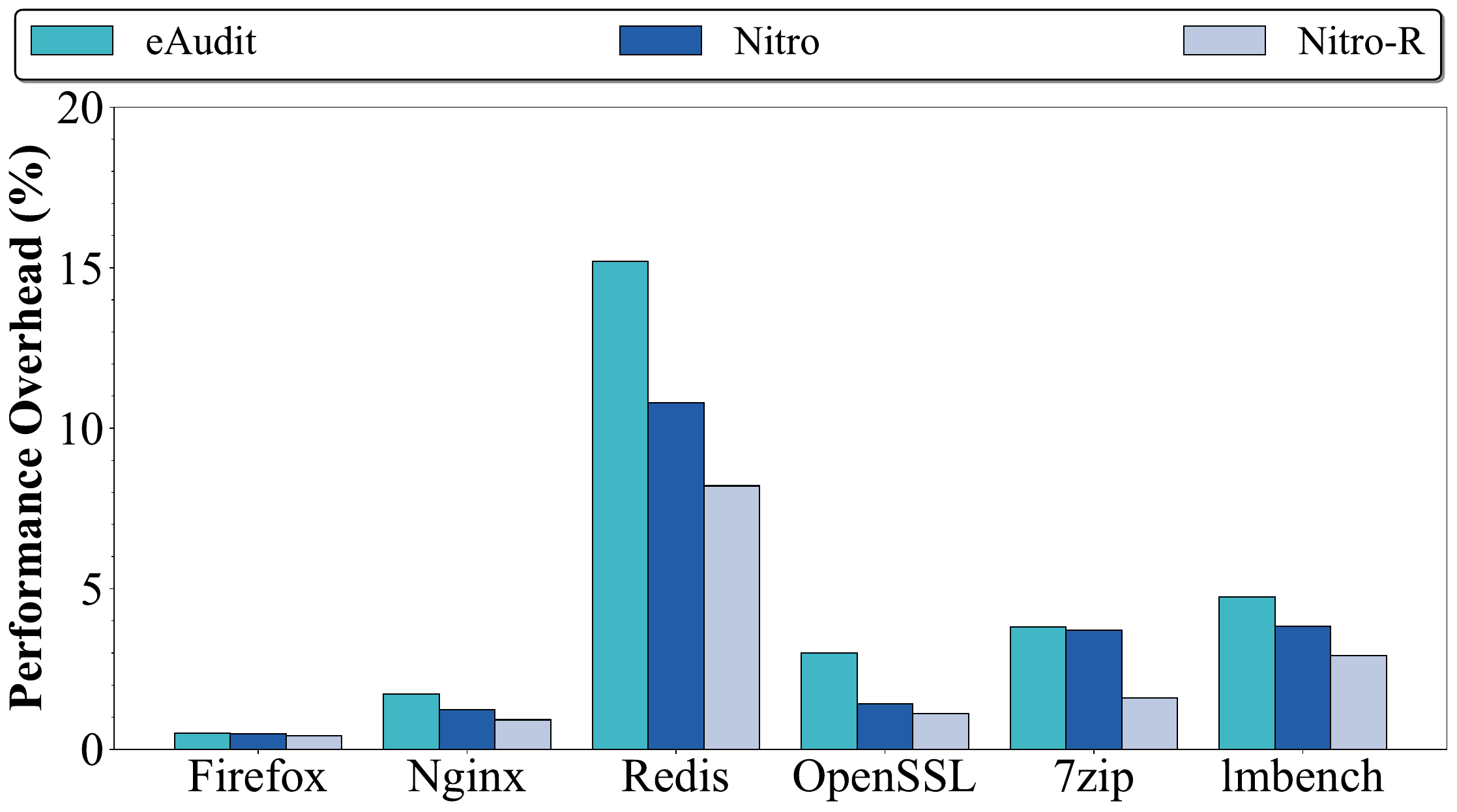}
    \caption{Runtime overhead comparison of \ed, \Sys, and \SysR under real-world benchmarks.}
    \label{fig:rw_sysR}
    \vspace{-2ex}
\end{figure}

\section{Remaning Experiment Results of \SysR} \label{D:sr_cmp}

\begin{figure}[t!]
    \centering
    \includegraphics[width=0.45\textwidth]{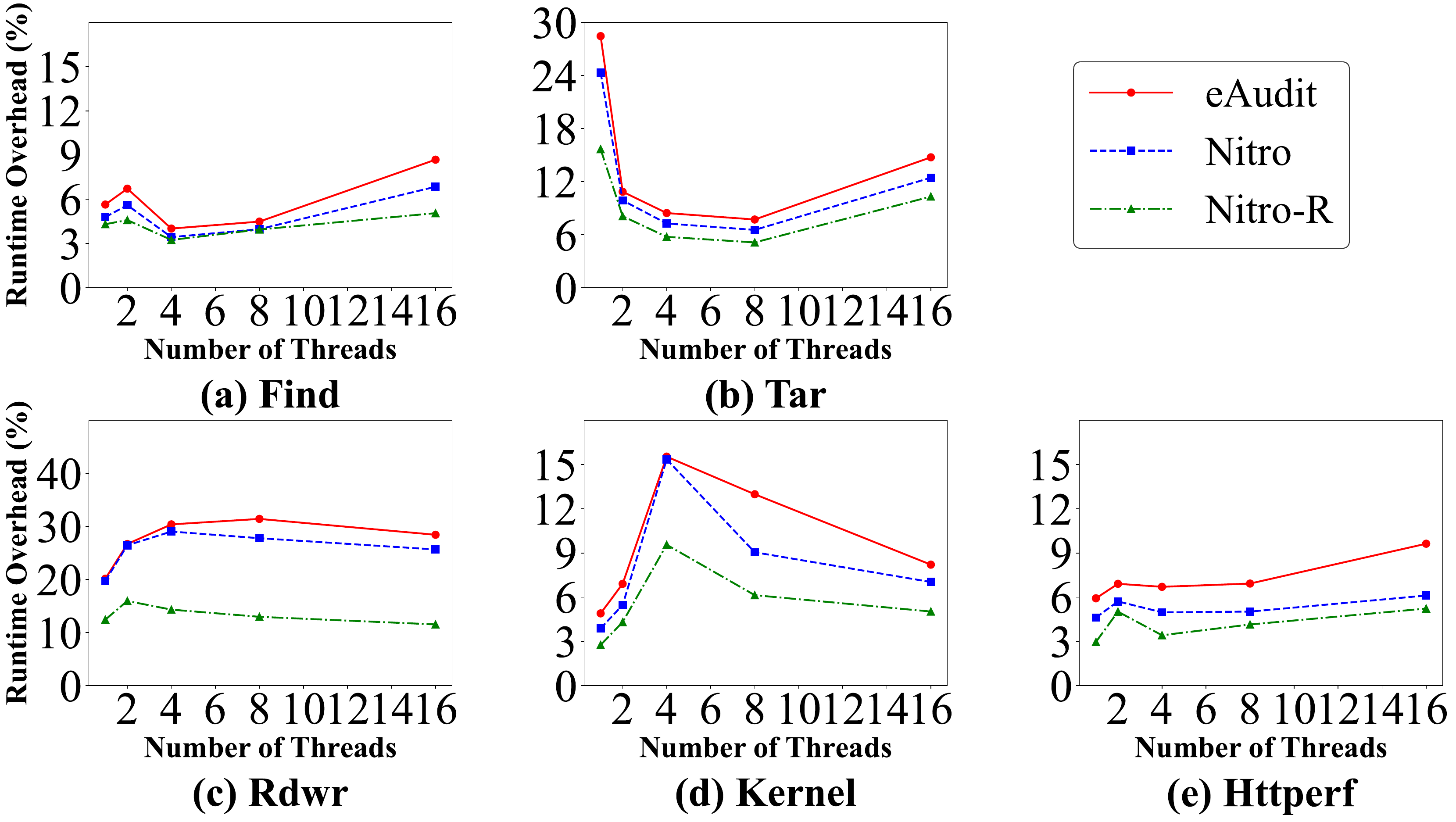}
    \caption{Remaining results for runtime overhead of \ed~\cite{Sekar2023eAA}, \Sys, and \SysR.}
    \label{fig:sysR_cmp_apd}
    \vspace{-1ex}
\end{figure}

Figure~\ref{fig:sysR_cmp_apd} shows the remainder of the seven stress-test benchmarks for comparing for comparing \SysR with \Sys and \ed~\cite{Sekar2023eAA}. Figure~\ref{fig:rw_sysR} presents the results of real-world benchmarks.
The trend here is the same as reported in \secref{s:eval}. 
Across both stress-test and real-world workloads, the overhead of \SysR remains within 1.2$\times$ of \Sys on average, and is significantly lower than that of \ed. For example, in the \texttt{tar} and \texttt{find} benchmarks, \SysR incurs only marginal overhead beyond \Sys despite additional signature generation, while still outperforming \ed by a wide margin.

This efficiency stems from \SysR's use of a kernel-level redundant log reduction mechanism implemented using eBPF’s built-in LRU map. Although the log reduction process introduces modest overhead during logging, it eliminates substantial downstream cost by suppressing unnecessary log entries before they reach the MAC and storage stages. In effect, this tradeoff results in net performance gains.
In particular, log reduction reduces the volume of logs transmitted to both the Per-CPU Array and the ring buffer, thereby lowering I/O pressure and minimizing memory copy overhead. It also decreases the size of the MAC input, leading to reduced cryptographic computation during signature generation. As a result, \SysR achieves strong runtime integrity protection while maintaining overhead even lower than \Sys.

\end{appendix}

\end{document}